# Pair-wise Markov Random Fields Applied to the Design of Low Complexity MIMO Detectors[1]

Seokhyun Yoon, *member*, *IEEE*, and Jun Heo, *member*, *IEEE*

**Abstract**: Pair-wise Markov random fields (MRF) are considered for application to the development of low complexity, iterative MIMO detection. Specifically, we consider two types of MRF, namely, the fully-connected and ring-type. For the edge potentials, we use the bivariate Gaussian function obtained by marginalizing the posterior joint probability density under the Gaussian assumption. Since the corresponding factor graphs are sparse, in the sense that the number of edges connected to a factor node (edge degree) is only 2, the computations are much easier than that of ML, which is similar to the belief propagation (BP), or sum-product, algorithm that is run over the fully connected factor graph. The BER performances for non-Gaussian input are evaluated via simulation, and the results show the validity of the proposed algorithms. We also customize the algorithm for Gaussian input to obtain the Gaussian BP that is run over the two MRF and proves its convergence in mean to the linear MMSE estimates. The result lies on the same line of those in [16] and [24], but with differences in its graphical model and the message passing rule. Since the MAP estimator for the Gaussian input is equivalent to the linear MMSE estimator, it shows the optimality, in mean, of the scheme for Gaussian input.

## I. INTRODUCTION

Recent works on multi-input and multi-output (MIMO) detection were mainly focused on the so-called sphere decoding [1-6]. Sphere decoding is a two stage detector in which the channel matrix is first converted into an upper triangular form and, utilizing this structure, a tree search is used for joint data detection. Since the full tree search has the same complexity of maximum likelihood (ML) detection, a sort of reduced search algorithm is applied by limiting the search space, e.g., the number of candidate symbols or radius at each tree search stage. One advantage of sphere decoding is that it can provide a tradeoff between performance and complexity by choosing an appropriate value of radius or list size. The performance of sphere decoding was shown to be quite close to that of ML with reasonable complexity [6].

---

[1] This work was supported by Basic Science Research Program through the National Research Foundation of Korea (NRF) funded by the Ministry of Education, Science and Technology (2009-0064557, 2010-0015582) This work was partially submitted to ICC 2011

Another type of MIMO detectors, which has not received much attention, is the channel truncation approach [7-10]. It is also a two stage detector, where the channel is first converted into a bi-diagonal or, more generally, a poly-diagonal form [9,10] and, utilizing the effective channel structure, a trellis search, e.g., Viterbi algorithm or the forward-backward algorithm [11][12], is used for post-joint detection. The scheme is similar to the concatenated channel shortening equalizer and maximum likelihood sequence estimator (MLSE) for the inter-symbol interference channel [13]. By employing channel shortening, rather than channel inversion, the noise enhancement that severely affects the performance can be eased, while the number of interferences is limited so that MLSE can be implemented with less complexity.

Graph based detection [14-20] is another class that is worthy of attention. The approaches are based on the belief propagation (BP) algorithm [21,22] that also has been extensively studied for decoding of the channel codes, such as the turbo codes and low density parity check codes. In these approaches, the MIMO channel is modeled as a fully-connected factor graph, which consists of a multiple $N$ factor nodes representing the received signal, a multiple $M$ variable nodes representing the hidden data, and the edges connecting the factor nodes with the variable nodes. The resulting graph has the maximal edge degree, i.e., every factor node is connected to every variable node. When applying the BP algorithm [21], or the sum-product algorithm [22], to such graphs, the complexity is as high as the ML or MAP detector due mainly to the marginalization operation required for the message update at the factor nodes. To reduce the computational complexity, the Gaussian BP has been considered in [16] and [17], where the input data and messages are all assumed to be Gaussian so that the message and posterior probability can be represented by a pair of mean and variance, resulting in a very simple message update rule. As shown in [16] and [17], however, the algorithm converges only to the linear minimum mean squared error (LMMSE) solution that is inferior to the ML detector for non-Gaussian input. On the other hand, in [18] and [20], complexity reduction via model simplification has been studied. Especially, in [18], it was suggested to prune some edges in the fully connected factor graph, based on the strength of the channel coefficients, in order to reduce the edge degree. By doing so, not only the number of messages is reduced, but also the marginalization operation on factor nodes can be performed at much less cost. Reduction in the marginalization cost is exponential with the edge degree reduction resulting in far less complexity than ML. The problem here, however, is that the performance loss can be more severe with the edge degree reduction.

Another interesting graph-based approach is the one in [20] based on the pair-wise Markov random field [23]. Here, noticing that BP may not work well for a loopy graph, a tree approximation was proposed on the basis of Kullback-Leibler distance (KLD) optimality criterion. One difference of this approach from other graph and BP based detectors is that it utilizes the Markov random field (MRF), rather than factor graphs. In MRF, we have only one type representing the hidden data and the edges reflecting the local dependency among them. The local dependency is represented by potential functions and, specifically in pair-wise MRF, they are functions of one or two variables. In fact, as noticed in [20] and [23] (also in [16] and [19]), a multivariate Gaussian function can be decomposed into a product of functions of one or two variables resulting in a fully connected pair-wise MRF.

In this paper, we examine two types of pair-wise MRF, i.e., the fully-connected and ring-type. However, instead of using the potential functions obtained from the direct decomposition of multivariate Gaussian function, we propose to use the potential functions that are obtained by marginalizing the posterior joint probability density under the Gaussian assumption. As will be shown later, the corresponding factor graph only has an edge degree of 2, as also implicated in [20] and [23], and the proposed scheme has much less complexity than that of ML/MAP. Unlike the one in [18], however, performance degradation is shown to be reasonable and, as shown in the simulation results, the bit error performance is shown to have close-to ML performance for non-Gaussian input.

This paper is organized as follows. In the next section, we briefly review the ML/MAP and the graph-based approach to MIMO detection. In Section III, the proposed iterative detection algorithm is derived for fully-connected and ring type pair-wise MRFs, respectively, for non-Gaussian input. In Section IV, assuming the Gaussian input, we customize the proposed algorithms for Gaussian input, i.e., Gaussian BP, and discuss its convergence property. The performances are evaluated and compared via link-level simulations in Section V and, finally, the concluding remarks are given in Section VI.

## II. MAP AND GRAPH-BASED DETECTION

In this section, we briefly review the channel model and some of the graph-based detectors related to our work.

*A. System model, MAP and graph-based detection*

A Gaussian MIMO system with a constant $N\times M$ channel matrix $\mathbf{H}$ ($N \geq M$) is modeled by

$$\mathbf{y} = \mathbf{Hx} + \mathbf{n} = \sum_{k=1}^{M} \mathbf{h}_k x_k + \mathbf{n} \tag{1}$$

where $\mathbf{x}$ is $M\times 1$ transmitted data symbol vector, $\mathbf{n}$ is $N\times 1$ noise vector, $\mathbf{y}$ is $N\times 1$ received signal vector and $\mathbf{h}_m$ is the $m$th column of $\mathbf{H}$. $\mathbf{n}$ is assumed to be complex Gaussian with mean $\mathbf{0}$ and covariance $E[\mathbf{nn}^H] = \sigma^2\mathbf{I}$ and the transmitted data symbol vector, $\mathbf{x}$, is assumed to have mean $\mathbf{0}$ and the covariance matrix $E[\mathbf{xx}^H] = \mathbf{I}$. In practice, each element of $\mathbf{x}$ is usually a $2^m$-ary symbol drawn from a finite alphabet set $\Xi$ of size $2^m$, such as QPSK or 16QAM.

*MAP detection*: The maximum *a posteriori* (MAP) detector selects $\mathbf{x}^*$ that maximizes the *a posteriori* likelihood

$$p(\mathbf{x} \mid \mathbf{y}) = \frac{p(\mathbf{y} \mid \mathbf{x}) p(\mathbf{x})}{p(\mathbf{y})}$$

where $$p(\mathbf{y} \mid \mathbf{x}) = \mathcal{CN}\left(\mathbf{y}; \mathbf{Hx}, \sigma^2\mathbf{I}\right) \tag{2}$$

$$p(\mathbf{x}) = \prod_{j=1}^{M} p(x_j) \tag{3}$$

with $\mathcal{CN}(\mathbf{y};\, \boldsymbol{\mu},\, \mathbf{C})$ representing the multivariate complex Gaussian PDF of mean $\boldsymbol{\mu}$ and covariance $\mathbf{C}$ defined as

$$\mathcal{CN}\left(\mathbf{y}; \boldsymbol{\mu}, \mathbf{C}\right) \equiv \frac{1}{(2\pi)^M \det \mathbf{C}} \exp\left( -\frac{1}{2} (\mathbf{y} - \boldsymbol{\mu})^H \mathbf{C}^{-1} (\mathbf{y} - \boldsymbol{\mu}) \right)$$

The search space of the MAP is an $M$-dimensional space, $\Xi^M$, and the complexity is $O(2^{mM})$. When using concatenated channel coding and MIMO, MIMO detector is required to produce soft-decision value, aka, log-likelihood ratio (LLR). Denoting the $j$th data symbol as $x_j(b_{j1}, b_{j2},\ldots, b_{jm})$, LLR of $b_{jk}$ can be obtained by first marginalizing $p(\mathbf{x}|\mathbf{y})$ over $\mathbf{x}\backslash x_j = (x_1, x_2,\ldots, x_{j-1}, x_{j+1},\ldots, x_M)$ to get

$$\begin{aligned} p(x_j = x \mid \mathbf{y}) &= \sum_{\mathbf{x}\backslash x_j \in \Xi^{M-1}} p(x_1, x_2, .., x_M \mid \mathbf{y}) \\ &= A \cdot \sum_{\mathbf{x}\backslash x_j \in \Xi^{M-1}} p(\mathbf{y} \mid x_1, x_2, .., x_j = x, ..., x_M) \prod_{k \neq j} p(x_k) \end{aligned} \tag{4}$$

where $A = p^{-1}(\mathbf{y})$ is a normalizing constant, and we used the assumption that $x_j$'s are independent of each other. The LLR for each bit is then computed as

$$LLR(b_{jk}) = \log\left( \frac{p(b_{jk} = 0 \mid \mathbf{y})}{p(b_{jk} = 1 \mid \mathbf{y})} \right) = \log\left( \frac{\sum_{\text{all } x_j : b_{jk} = 0} p(x_j \mid \mathbf{y})}{\sum_{\text{all } x_j : b_{jk} = 1} p(x_j \mid \mathbf{y})} \right) \tag{5}$$

The prohibitive complexity, especially when $m$ and $M$ are large, comes from the marginalization operation in (4) and (5).

Since the noise covariance is given by a diagonal matrix, the marginalization can also be performed for each element of **y**, i.e.,

$$\begin{aligned} p(x_j = x | y_k) &= \sum_{\mathbf{x}\backslash x_j \in \Xi^{M-1}} p(x_1, x_2, .., x_M \mid y_k) \\ &\propto p(x_j) \sum_{\mathbf{x}\backslash x_j \in \Xi^{M-1}} p(y_k \mid x_1, x_2, .., x_j = x, ..., x_M) \prod_{k \neq j} p(x_k) \end{aligned} \quad (6)$$

where

$$p(y_k \mid x_1, x_2, .., x_j = x, ..., x_M) = p(y_k \mid \mathbf{x}) = \mathcal{CN}\left(y_k ; \sum_{j=1}^{M} h_{kj} x_j, \sigma^2\right) \quad (7)$$

such that $p(\mathbf{y}|\mathbf{x}) = \Pi_k\ p(y_k|\mathbf{x})$. Note that the marginalization in (6) as well as in (4) is performed over $M$−1 dimensional space. Since the number of lattice points in $M$−1 dimensional space is $2^{m(M-1)}$ and the marginalization must be performed for the total number $2^m$ states of $x_j$, the complexity remains the same as that of MAP.

*Graph based detection (BP over fully-connected factor graph)*: The MAP detection in (4) or (6) is very useful for turbo equalization [27][28], where the reader can find a vast amount of literature showing the validity of iterative MIMO detection and channel decoding. Although the turbo equalization is not our focus in this paper, it is worthy of paying attention to the iterative detection as in [18], i.e., the BP over the fully-connected factor graph. As a matter of fact, the MAP detection in (4) can be regarded as a belief propagation that is run over the singly connected factor graph in Fig.1 (a), where each variable node, representing a data symbol, first passes *a priori* information to the factor node, representing the received vector, **y**. The factor node then provides each variable node with the corresponding *a posteriori* likelihood by computing the marginalization in (4). Since the graph is singly connected and all variable nodes are connected via one factor node, the BP over this graph will surely converge, in one iteration, to the correct *a posteriori* probability. The graph based detection in [18], on the other hand, is a BP over the fully connected factor graph in Fig.1 (b), which can be summarized as follows.

BP 1 over the fully-connected factor graph [18]

(1) Initialize, for all edge,

$$\lambda_{j\to i}(x_j) = p(x_j) \quad (8)$$

(2) Factor Node computation:

$$\pi_{i\to j}(x_j = x) = A\cdot \sum_{\mathbf{x}\backslash x_j \in \Xi^{M-1}} p(y_{\mathrm{k}} \mid x_1, x_2,.., x_j = x,..., x_M)\prod\nolimits_{k\neq j} \lambda_{k\to i}(x_k) \quad (9)$$

(3) Belief update:

$$b(x_j) = p(x_j)\cdot \prod\nolimits_{k=1}^{M} \pi_{k\to j}(x_j) \quad (10)$$

(4) Variable Node computation:

$$\lambda_{j\to i}(x_j) = p(x_j)\cdot \prod_{k\neq i} \pi_{k\to j}(x_j) = b(x_j)/\pi_{i\to j}(x_j) \quad (11)$$

with $p(y_{\mathrm{k}} \mid x_1, x_2,.., x_j = x,..., x_M)$ given by (7). The message update (9)-(11) are repeated by a pre-defined number or until the belief does not change any more.

Although its complexity is similar to the MAP detector, it provides us a base structure for the development of low complexity detector as to be discussed in the next.

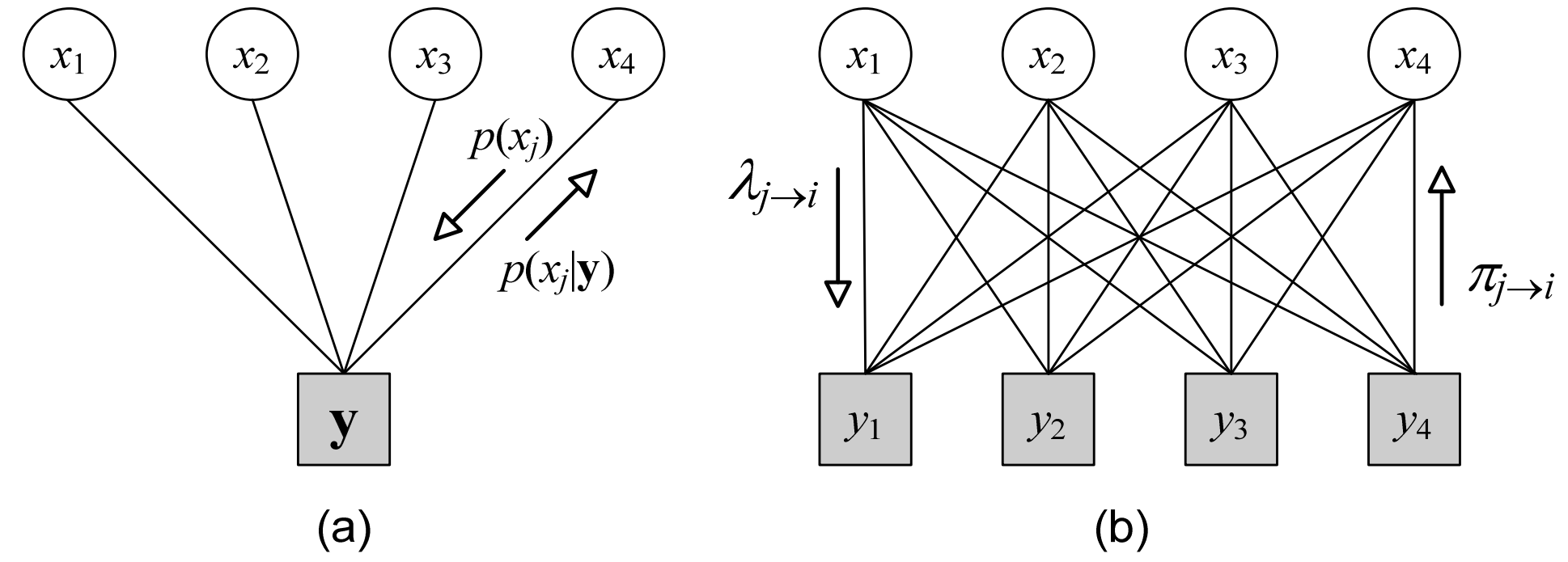

Fig.1 Factor graphs for a 4x4 MIMO channel

*B. Complexity reduction via edge pruning*

To reduce the computational burden of the marginalization in (9) for non-Gaussian input, [18] proposed to prune some edges for which the corresponding variable and observation nodes are weakly coupled together, e.g., those variable-factor node pairs with small value of $|h_{jk}|$. By using only $d_f < M$ edges per factor node (i.e., pruning $M-d_f$ edges), the complexity is reduced by a factor of $1/2^{m(M-d_f)}$ relative to ML/MAP or the BP 1 of complexity $O(2^{mM})$. The problem of this scheme is that $d_f$ must not be too small to ensure reasonable performance as shown in [18].

An alternative strategy to prune edges has been proposed in [7-10], where the edges are pruned by a linear transformation, namely, the poly-diagonalization. The simplest one is the bi-diagonalization, where the effective channel is, for example, given by

$$\mathbf{H}_{eff} = \begin{bmatrix} a_{11} & 0 & 0 & a_{14} \\ a_{21} & a_{22} & 0 & 0 \\ 0 & a_{32} & a_{33} & 0 \\ 0 & 0 & a_{43} & a_{44} \end{bmatrix} \tag{12}$$

for $M = 4$. The resulting channel structure looks like a "tail-biting" ISI channel and, hence, utilizing the poly-diagonal structure of the underlying channel matrix, a tail-biting trellis decoding (known also as the forward-backward algorithm) is applied to produce the a posterior likelihoods. In this approach, the edge degree, $d_f$, could be as small as 2 for a reasonable performance degradation.

The idea in the poly-dagonalization procedure is to use a multi-modal linear estimator, $\mathbf{c}_{\Phi(j)}$ for $j = 1,2,\ldots,M$, either in zero-forcing (ZF) sense or in minimum mean squared error (MMSE) sense, where $\Phi(j)$ is the indices set consisting of ordered numbers $\{(j-l)_M+1;\ l=1,\ldots,L\}$ with $L$ defined as the order of poly-diagonalization. The index $j$ in $\Phi(j)$ corresponds to the target signal, all others in $\Phi(j)$ are its companion, while those not in $\Phi(j)$ are considered as interference to be nullified or suppressed. For example, for the case $M = 4$ and $L = 2$, $\Phi(1) = \{4, 1\}$ and $\Phi(3) = \{2, 3\}$. By applying $\mathbf{c}_{\Phi(j)}$ to the received vector $\mathbf{y}$, we have

$$\mathbf{c}_{\Phi(j)}^H \mathbf{y} = \underbrace{\sum_{k\in\Phi(j)} \mathbf{c}_{\Phi(j)}^H \mathbf{h}_k x_k}_{\text{signals}} + \underbrace{\sum_{k\in\overline{\Phi}(j)} \mathbf{c}_{\Phi(j)}^H \mathbf{h}_k x_k}_{\text{interference to suppress}} + \mathbf{c}_{\Phi(j)}^H \mathbf{n} \tag{13}$$

where $\overline{\Phi}(j)$ is the complementary set of $\Phi(j)$. $\mathbf{c}_{\Phi(j)}$ is chosen such that the interference in the r.h.s. of (13) is nullified (in ZF sense) or the interference plus noise is suppressed (in MMSE sense). In MMSE sense, the multi-modal MMSE estimator, $\mathbf{c}_{\Phi(j)}$, can be chosen such that for arbitrary constant $B$

$$\mathbf{c}_{\Phi(j)} = \arg\max_{\mathbf{c}\in C^N} \frac{\mathbf{c}^H \mathbf{h}_j \mathbf{h}_j^H \mathbf{c}}{\mathbf{c}\mathbf{K}_{\Phi(j)}\mathbf{c}} = B \cdot \mathbf{K}_{\Phi(j)}^{-1} \mathbf{h}_j \tag{14}$$

where $\mathbf{K}_{\Phi(j)}$ is the partial covariance matrix of the noise plus nominal interferences, $\{\mathbf{h}_k; k \in \overline{\Phi}(j)\}$, given by

$$\mathbf{K}_{\Phi(j)} = \sigma^2 \mathbf{I} + \sum\nolimits_{k\in\overline{\Phi}(j)} \mathbf{h}_k \mathbf{h}_k^H$$

For convenience, we set $B = 1$. Then, with the choice in (14), (13) can be rewritten as

$$\mathbf{c}_{\Phi(j)}^H \mathbf{y} = y'_j = \underbrace{\sum_{k\in\Phi(j)} a_{j,k} x_k}_{\text{signals to enhance}} + n'_j \tag{15}$$

where $n'_j$ is the suppressed interference plus noise and

$$a_{j,k} = \mathbf{c}_{\Phi(j)}\mathbf{h}_k = \mathbf{h}_j^H \mathbf{K}_{\Phi(j)}^{-1}\mathbf{h}_k \tag{16}$$

$$E\,|\,n'_j\,|^2 = \mathbf{c}_{\Phi(j)}^H \left(\sigma^2\mathbf{I} + \sum\nolimits_{k\in\overline{\Phi}(j)} \mathbf{h}_k\mathbf{h}_k^H\right)\mathbf{c}_{\Phi(j)} = \mathbf{c}_{\Phi(j)}^H\mathbf{h}_j \equiv \sigma_j^2 \tag{17}$$

Based on the Gaussian approximation of $n'_j$, we have

$$p(y'_j \mid x_{j-L}, x_{j-L+1}, \ldots, x_j) \approx \mathcal{CN}\left(y'_j; \sum\nolimits_{k\in\Phi(j)} a_{j,k}x_k, \sigma_j^2\right) \tag{18}$$

and the *a posterior* likelihood, corresponding to (6), is obtained by marginalizing (18) as

$$p(x_j = x \,|\, y'_j) \propto \sum_{(x_{j-L+1},\ldots,x_{j-1})\in\Xi^{L-1}} p(y'_j \mid x_{j-L+1}, x_{j-L+2}, \ldots, x_j) \prod\nolimits_{k\in\Phi(j)} p(x_k) \tag{19}$$

In (18) and (19), we omitted the modulo operation $(\cdot)_M$ for notational convenience. Since the marginalization in (19) is performed over only $L-1$ dimensional space and typically $L = 2$ or 3, the computational complexity can be far less than that in (6).

Defining the matrix $\mathbf{C}$ as a $N\times M$ matrix with its $j$th column given by $\mathbf{c}_{\Phi(j)}$, the effective channel model is given by

$$\mathbf{C}^H\mathbf{y} = \mathbf{H}_{eff}\mathbf{x} + \mathbf{n}' \tag{20}$$

where $\mathbf{n}' = [n_1', n_2', \ldots, n_M']^{\mathrm{T}}$ and the effective channel, $\mathbf{H}_{eff}$, is given by (12). Fig.2 (a) shows the graphical model for the poly-diagonalized MIMO channel with $M = 4$ and $L = 2$ and Fig.2 (b) is the same model plotted as a ring, with which the implementation of belief propagation algorithm is clearer. With $L = 2$, (18) and (19) become

$$p(y'_j \mid x_{(j-1)}, x_j) = \mathcal{CN}\left(y'_j; a_{j,j}x_j + a_{j,j-1}x_j, \sigma_j^2\right) \tag{21}$$

$$p(x_j = x \,|\, y'_j) \propto p(x_j)\cdot \sum_{x_{j-1}\in\Xi} p(y'_j \mid x_{j-1}, x_j)\,p(x_{j-1}) \tag{22}$$

where (22) is the factor node message passing in the forward direction. Replacing $p(x_{j-1})$ and $p(x_j|y_j')$ with the incoming message, $\lambda_{j,j-1}(x_{j-1})$, and the outgoing message, $\pi_{j,j-1}(x_j)$, respectively, the message flow from the variable node $j-1$ to $j$ becomes

$$\pi_{j,j-1}(x_j \,|\, y'_j) \propto p(x_j)\cdot \sum_{x_{j-1}\in\Xi} p(y'_j \mid x_{j-1}, x_j)\cdot \lambda_{j,j-1}(x_{j-1}) \tag{23}$$

The backward message passing can also be obtained in a similar way, i.e., using (21), we also have

$$p(x_{j-1} = x \,|\, y'_j) \propto p(x_{j-1})\cdot \sum_{x_{j-1}\in\Xi} p(y'_j \mid x_{j-1}, x_j)\,p(x_j) \tag{24}$$

And, by replacing $p(x_j)$ and $p(x_{j-1} \mid y_j')$ with $\lambda_{j-1,j}(x_j)$ and $\pi_{j-1,j}(x_{j-1})$, respectively, we have

$$\pi_{j-1,j}(x_{j-1} \,|\, y'_j) \propto p(x_{j-1})\cdot \sum_{x_j\in\Xi} p(y'_j \mid x_{j-1}, x_j)\cdot \lambda_{j-1,j}(x_j) \tag{25}$$

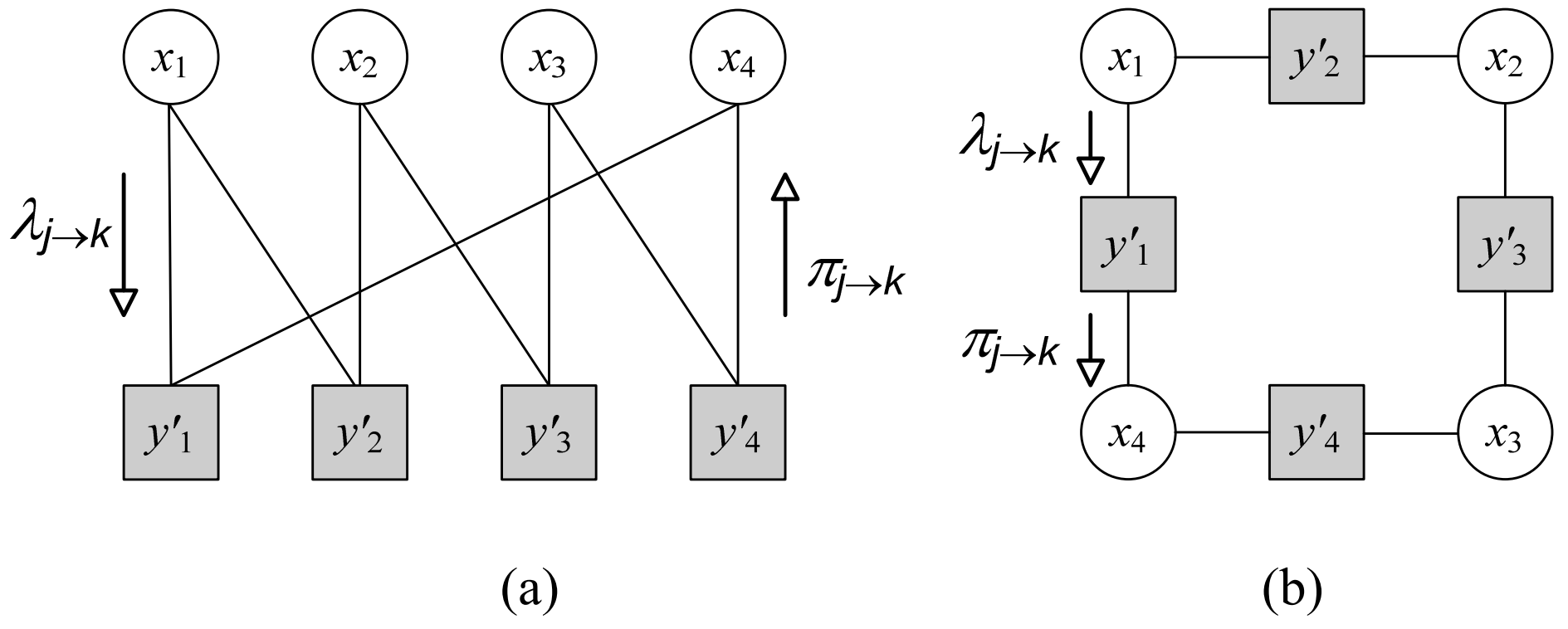

Fig.2 Factor graphs for a 4x4 bi-diagonalized MIMO channel

The detection algorithm in (21)-(25) is equivalent to the graph-based, iterative detection for the ISI channel in [26][27] for a channel length of 2. As in [26], this message passing detection can also be easily extended to more general cases where $L$>2. As mentioned in [9], the direct graphical model has many local loops and the problem become complicated. To develop the corresponding BP rule, which is intuitively more meaningful for the general case of $L$>2, we define the "state" as a group of neighboring symbols. By deriving the BP rule among these state variables, the forward and backward message passing principle can also be applied for $L$ >2 [9].

*D. Sphere decoding*

Sphere decoding can also be described in the context of graph-based detection. In sphere decoding, the QR decomposition is first used to effectively triangularize the channel, and a tree search is then applied for post joint detection. In QR decomposition, the channel matrix, $\mathbf{H}$, is decomposed into $\mathbf{H} = \mathbf{QR}$, where $\mathbf{Q}$ is $N{\times}N$ unitary matrix such that $\mathbf{Q}^{\mathrm{H}}\mathbf{Q} = \mathbf{I}$ and $\mathbf{R}$ is an upper triangular matrix. By pre-multiplying the received signal vector $\mathbf{y}$ with $\mathbf{Q}^{H}$, we obtain the effective channel given by $\mathbf{y}' = \mathbf{Q}^{\mathrm{H}}\mathbf{y} = \mathbf{Rx} + \mathbf{n}'$ where $\mathbf{n}' = \mathbf{Q}^{\mathrm{H}}\mathbf{n}$ for which the covariance matrix is $E[\mathbf{n}'\mathbf{n}'^{H}] = \sigma^2\mathbf{I}$. Since the preprocessor $\mathbf{Q}$ is a unitary matrix that does not alter the noise correlation or singular values, there is no loss in capacity by pre-multiplying $\mathbf{Q}$ with the received signal vector $\mathbf{y}$. Fig.3 shows an example graphical model for the triangularized MIMO channel.

Once the channel is triangularized, the sphere decoding employs a sort of reduced tree search, where, starting from the root node, the algorithm searches only those nodes that lie within the sphere of radius, $r$, centered at the received signal vector $\mathbf{y}'$. Apparently, the sphere decoding is equivalent to the ML detector when using a full tree search. And, certainly, there exists a tradeoff between the performance and complexity that can be controlled by the radius, $r$. That is, with a small $r$, the number of nodes to be searched along the tree can be considerably reduced, while it cannot be too small since the set of candidate nodes may be dried out and/or the nearest lattice may be excluded before we reach the final tree stage resulting in inaccurate detection.

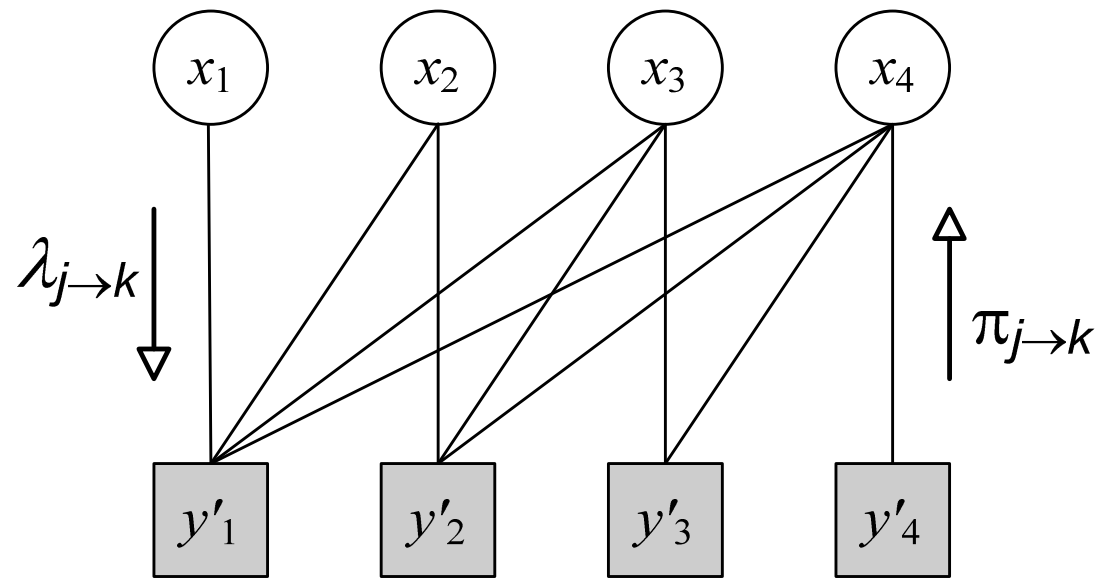

Fig.3 Factor graph for a 4x4 triangular MIMO channel

## III. Detection Algorithm based on Pair-wise Markov Random Field

In this section, we develop low complexity iterative MIMO detection algorithms based on the pair-wise Markov random fields. We consider two types, namely, a fully-connected MRF and a ring-type-MRF, and derive the corresponding BP algorithms (sum-product algorithm) that work for the non-Gaussian input. As will be shown later, BP over the ring-type MRF is effectively equivalent to the one in [10] with a slight difference.

### *A. Detection algorithm based on fully-connected MRF*

MRF is an undirected graph that describes local dependencies among a set of random variables. In MRF, the joint PDF of all r.v. involved can be represented by a product of joint PDF of each clique. The pair-wise MRF [23] is defined as MRF where, for each clique, the joint PDF of a set of random variables can be represented by a product of two dimensional joint PDFs of any two neighbors. Let $N = \{1,2,\ldots,M\}$ be the set of nodes in the MRF corresponding to the random variables $x_1, x_2,\ldots,x_M$, respectively, and let $E$ be the set of all edges connecting these nodes. For a compact expression, we also denote the edge connecting nodes $j$ and $k$ as $e(j,k)$ and the set of neighbors of the $j$th node as $N(j)$. In pair-wise MRF, the a posteriori joint PDF $p(x_1, x_2,\ldots,x_M|\mathbf{y})$ is modeled by a product of pair-wise potential functions [23][17], e.g.,

$$\hat{p}(x_1, x_2, \ldots x_M \mid \mathbf{y}) = A \prod_{i \in N} \psi_i(x_i) \prod_{(i,j):e(i,j) \in E} \phi_{ij}(x_i, x_j) \tag{26}$$

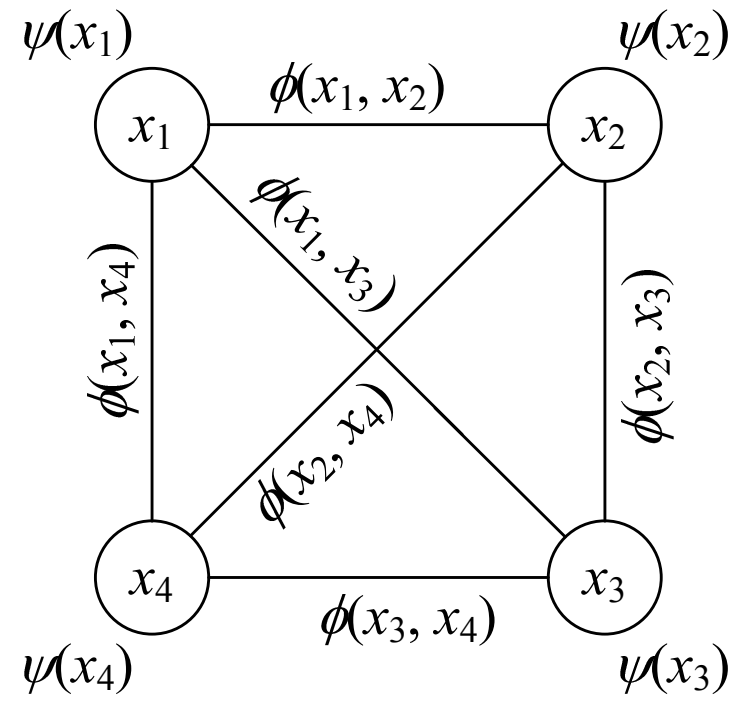

Fig.4 Fully-connected Markov Random Fields with 4 variables

Fig.4 shows a MRF with 4 variables, where self-potential, $\psi(x_i)$, is assigned to each node and the edge potential, $\phi(x_i, x_j)$ is assigned to each edge. One choice of such potential function can be, as suggested in [16], [17] and [20],

$$\phi_{ij}(x_i, x_j) = A_{ij} \exp\left( -\frac{1}{\sigma^2} \mathrm{Re}[x_i^* R_{ij} x_j] \right) \tag{27}$$

$$\psi_i(x_i) = A_i \exp\left( -\frac{1}{\sigma^2} \mathrm{Re}[x_i^* y_j' - R_{ii} \mid x_i \mid^2] \right) \tag{28}$$

where $R_{ij} = \mathbf{h}_i^H \mathbf{h}_j$ and $y_j' = \mathbf{h}_i^H \mathbf{y}$. Denoting the (incoming) message from the $i$th node to the $j$th as $\pi_{i \to j}(x_j)$, the belief propagation through the pair-wise MRF can be described as [17]

$$\pi_{i \to j}(x_j) = \alpha \sum_{x_i \in \Xi} \psi_i(x_i) \phi_{ij}(x_i, x_j) \prod_{k \in N(i) \setminus j} \pi_{k \to i}(x_i) \tag{29}$$

where $N(i)\setminus j$ is the set of neighbors of node $i$ excluding node $j$. In (29), the incoming messages are combined first to produce the extrinsic information, $\Pi_{k \in N(i) \setminus j} \pi_{k \to i}(x_i)$, and they are then "translated" by the potential function, $\psi_i(x_i)\phi_{ij}(x_i,x_j)$. The belief on the variable, $x_j$, is given by

$$b(x_j) = \prod_{k \in N(j)} \pi_{k \to j}(x_j) \tag{30}$$

Note that, when $x_j$'s are Gaussian (continuous) random variables, the messages, $\pi_{i \to j}(x_j)$, can be summarized by a pair of mean and variance giving us a very simple message update rule as in [16] and [17]. As pointed out in [17] and proved in [16], however, the Gaussian belief propagation based on the above potential functions results in the solution of nothing but the linear MMSE estimator. As discussed before, it is optimum for the Gaussian input in the sense that it is the true *a posterior* probability. However, for non-Gaussian input, it is generally inferior to the ML/MAP detector.

In this paper, we propose to use the pair-wise MRF but with different choice and usage of the edge potentials. Specifically, by assigning to each edge $e(i, j)$ a pair-wise joint PDF, $p(x_i, x_j | \mathbf{y})$, derived from the *a posteriori* PDF, $p(x_1, x_2, \ldots, x_M | \mathbf{y})$, under Gaussian assumption, we model the component-wise *a posteriori* probability, the $j$th, as a product of these edge potentials, i.e.,

$$\begin{aligned} \hat{p}(x_j | \mathbf{y}) &= A_j \prod_{i \in N(j)} \left( \sum_{x_i} p(x_i, x_j | \mathbf{y}) \right) \\ &= A_j \prod_{i \in N(j)} \left( \sum_{x_i} p(x_j | x_i, \mathbf{y}) p(x_i | \mathbf{y}) \right) \end{aligned} \tag{31}$$

where $A_j$ is the normalizing constant. Note that $\hat{p}(x_i | \mathbf{y})$ is certainly not equal to the true component wise *a posteriori* probability $p(x_i | \mathbf{y})$ obtained by (4). However, by replacing $p(x_i|\mathbf{y})$ in the summand with the estimates of its neighbors, $\hat{p}(x_i | \mathbf{y})$, we obtain a set of recursive relations giving us a desired message passing rule, where the message is given by $\hat{p}(x_i | \mathbf{y})$ and the translation is performed by $p(x_j | x_j, \mathbf{y})$.

Now, we take a look at the edge potential, $p(x_j, x_j | \mathbf{y})$, and the translation function, $p(x_j | x_j, \mathbf{y})$, derived from it. They are related by

$$p(x_i, x_j | \mathbf{y}) p(\mathbf{y}) = p(\mathbf{y} | x_i, x_j) p(x_i, x_j) = p(x_j | x_i, \mathbf{y}) p(x_i, \mathbf{y})$$

resulting in

$$p(x_j | x_i, \mathbf{y}) = \frac{p(\mathbf{y} | x_i, x_j) p(x_j)}{p(\mathbf{y} | x_i)} \tag{32}$$

where we used (3). For non-Gaussian input, it is quite difficult to derive $p(\mathbf{y}| x_i, x_j)$ and $p(\mathbf{y}|x_j)$ in the right hand side of (32). Hence, at this step, we assume $x_j$'s are Gaussian, even though the messages will not be treated as such. Under the Gaussian assumption on $x_j$, each term in (32) is given by

$$p(\mathbf{y} | x_i, x_j) = \mathcal{CN}\left(\mathbf{y}; \mathbf{h}_i x_i + \mathbf{h}_j x_j, \mathbf{K}_{\{j,i\}}\right) \tag{33}$$

$$p(\mathbf{y} | x_i) = \mathcal{CN}\left(\mathbf{y}; \mathbf{h}_i x_i, \mathbf{K}_{\{i\}}\right) \tag{34}$$

where

$$\begin{aligned} \mathbf{K}_\Phi &= \sigma^2 \mathbf{I} + \sum\nolimits_{k \notin \Phi} \mathbf{h}_k \mathbf{h}_k^H \\ &= \sigma^2 \mathbf{I} + \sum\nolimits_{k=1}^{M} \mathbf{h}_k \mathbf{h}_k^H - \sum\nolimits_{k \in \Phi} \mathbf{h}_k \mathbf{h}_k^H \end{aligned} \tag{35}$$

for $\Phi = \{i, j\}$ or $\{i\}$. Moreover, the Gaussian assumption leads us to a much simpler expression. That is, from the information theoretic optimality (sufficiency) of the MMSE estimator for Gaussian input, we have

$$I(x_j;\mathbf{y}\,|\,x_i) = I(x_j;\mathbf{c}_{j|i}^H\mathbf{y}\,|\,x_i) \tag{36}$$

where

$$\mathbf{c}_{j|i} = \mathbf{K}_{\{j,i\}}^{-1}\mathbf{h}_j\,. \tag{37}$$

is the conditional MMSE estimator for $x_j$ given $x_i$. Denoting $y'_{ji} = \mathbf{c}_{j|i}^H\mathbf{y}$, we have, in a similar way to (13)-(15),

$$y'_{j|i} = \mathbf{c}_{j|i}^H\mathbf{y} = a_{j|i,j}x_j + a_{j|i,i}x_i + n'_{j|i} \tag{38}$$

where

$$a_{j|i,k} = \mathbf{c}_{j|i}^H\mathbf{h}_k = \mathbf{h}_j^H\mathbf{K}_{\{j,i\}}^{-1}\mathbf{h}_k \quad \text{for } k = i \text{ or } j \tag{39}$$

$$E\,|\,n'_{j|i}\,|^2 = \mathbf{c}_{j|i}^H\mathbf{K}_{\{j,i\}}\mathbf{c}_{j|i} = \mathbf{h}_j^H\mathbf{K}_{\{j,i\}}^{-1}\mathbf{h}_j \equiv \sigma_{j|i}^2 \tag{40}$$

The sufficiency in (36) suggests using the following as the translation function, instead of (32), i.e.,

$$p(x_j\,|\,x_i, y'_{j|i}) = \frac{p(y'_{j|i}|x_i, x_j)\,p(x_j)}{p(y'_{j|i}|x_i)} \tag{41}$$

with

$$p(y'_{j|i}|x_i, x_j) = \mathcal{CN}(y'_{j|i}; a_{j|i,j}x_j + a_{j|i,i}x_i, \sigma_{j|i}^2) \tag{42}$$

$$p(y'_{j|i}|x_i) = \mathcal{CN}(y'; a_{j|i,i}x_i, \sigma_{j|i}^2 + |\,a_{j|i,j}\,|^2) \tag{43}$$

In (42) and (43), we used the Gaussian assumption, $p(x_j) = \mathcal{CN}(x_j;0,1)$. Plugging (42) and (43) into (41) and by replacing $p(x_j)$ with $\mathcal{CN}(x_j;0,1)$, we have the simplified translation function from the derivation in the appendix.

$$\begin{aligned} p(x_j|x_i, y'_{j|i}) &= \mathcal{CN}\left(x_j; \frac{a_{j|i,j}^*}{\sigma_{j|i}^2 + |\,a_{j|i,j}\,|^2}\left(y'_{j|i} - a_{j|i,i}x_i\right), \frac{\sigma_{j|i}^2}{\sigma_{j|i}^2 + |\,a_{j|i,j}\,|^2}\right) \\ &= \mathcal{CN}\left(x_j; \frac{1}{1+\sigma_{j|i}^2}\left(y'_{j|i} - a_{j|i,i}x_i\right), \frac{1}{1+\sigma_{j|i}^2}\right) \end{aligned} \tag{44}$$

where, in the last line, we used the fact that $a_{j|i,j}$ is real valued and is equal to $\sigma^2_{j|i}$. Note that in (44), the mean is the conditional MMSE estimate of $x_j$ given $x_i$.

Using the equations from (37) to (44), the message passing rule in (31) can be simplified and summarized as follows.

BP 2 over the fully-connected MRF

Given the messages in the previous iteration, $\pi_{k\to i}(x_i)$

(1) Compute the extrinsic information for all pairs $(i,j)$ with $i\neq j$

$$\lambda_{i\to j}(x_i)=\prod_{k\in N(i)\backslash j}\pi_{k\to i}(x_i) \qquad (45)$$

(2) Translate the message $\lambda_{i\to j}(x_i)$ to $\pi_{i\to j}(x_j)$

$$\pi_{i\to j}(x_j)=\sum_{x_i\in\Xi}p(x_j\mid x_i,y'_{j|i})\cdot\lambda_{i\to j}(x_i) \qquad (46)$$

with $p(x_j|x_i,y'_{j|i})$ given by (44). (45) and (46) are computed for all edges in both directions, and the message update is repeated by a pre-defined number or until the belief does not change any more and, finally, the belief is obtained the same as that in (30).

Note that, although we adopted the Gaussian input assumption to obtain the Gaussian translation function (44), the messages, $\pi_{i\to j}(x_j)$ and $\lambda_{i\to j}(x_i)$, in the above algorithm may not necessarily be Gaussian.

*B. Detection algorithm based on Ring-type-MRF*

It is also interesting to note that the message passing rule 2 is quite similar to the forward-backward recursion in [9] and [10], i.e., (23) and (25), with two differences. One is in the underlying structure and the other in message translation. To clarify the similarity and difference, we consider the ring-type MRF shown in Fig.5. In this ring-type MRF, each node has only two neighbors and, hence, (31) becomes

$$\hat{p}(x_j\mid\mathbf{y})=A_j\left(\sum_{x_{j+1}\in\Xi}p(x_j,x_{j+1}\mid\mathbf{y})\right)\left(\sum_{x_{j-1}\in\Xi}p(x_j,x_{j-1}\mid\mathbf{y})\right) \qquad (47)$$

Note that, in the computation of extrinsic information, the incoming message from one neighbor is simply passed to the other so that the message update rule (45) and (46) can be modified as follows:

BP 3 over the ring-type MRF (Forward-backward recursion)

Extrinsic information:

$$\lambda_{j\to j\pm1}(x_j) = \pi_{j\mp1\to j}(x_j) \tag{48}$$

Message translation:

$$\pi_{j\to j\pm1}(x_{j\pm1}) = \sum_{x_j\in\Xi} p(x_{j\pm1} \mid x_j, y'_{j|j\pm1}) \cdot \lambda_{j\to j\pm1}(x_j) \tag{49}$$

After a pre-defined number of iteration, the belief is obtained by

$$b(x_j) = \pi_{j+1\to j}(x_j) \cdot \pi_{j-1\to j}(x_j) \tag{50}$$

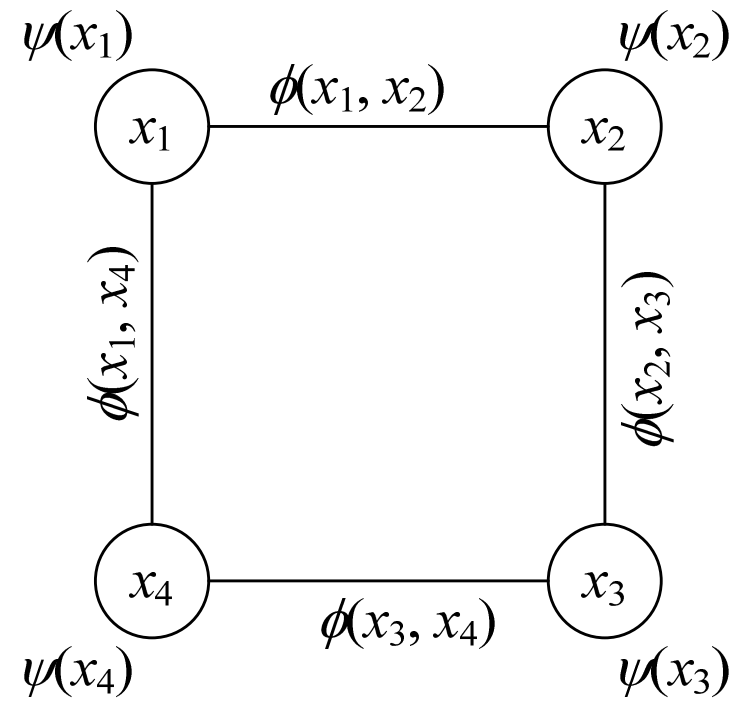

Fig.5 Ring-type Markov Random Fields with 4 variables

This message update rule is a forward-backward algorithm similar to (23) and (25), i.e., the message from the ($j$–1)th node to the $j$th node corresponds to the forward message, and the one from the ($j$+1)th node to the $j$th node corresponds to the backward message. The difference is in the message translation. In (49), the message translation from the $j$th node to the $i$th and its reverse utilize a different translation function, i.e.,

$$y'_{j|i} \neq y'_{i|j} \Rightarrow p(x_j \mid x_i, y'_{j|i}) = \frac{p(y'_{j|i} | x_i, x_j) p(x_j)}{p(y'_{j|i} | x_i)} \neq p(x_i \mid x_j, y'_{i|j}) \tag{51}$$

This means the branch metrics used for the forward and backward recursion are separately optimized to maximize their SINR, as also proposed in [10]. The translation function is also different from the branch metric in [10], i.e., the mean and variance in (44) have MMSE scaling by a factor of $a^*_{j|i,j} / (\sigma^2_{j|i} + |a_{j|i,j}|^2)$ and $1/(\sigma^2_{j|i} + |a_{j|i,j}|^2)$, instead of $a^*_{j|i,j} / |a_{j|i,j}|^2$ and $1/|a_{j|i,j}|^2$, though it has a minor impact on the error rate performances.

The factor graphs corresponding to the fully-connected MRF and the ring-type MRF are shown in Fig.6 (a) and (b), respectively, where the observation used for the message translation from the $j$th node to the $i$th and its reverse is clearly denoted by $y'_{j|i}$ and $y'_{i|j}$, respectively.

Note that, for ring-type MRF, we obtain different performance with a different antenna permutation, as also noted in [7, 8], while, in the fully-connected one, we do not need antenna permutation, which is one possible advantage of the latter to the former.

Since the proposed MRFs have short cycle(s), it is quite questionable whether or not the BP 2 and 3 will converge. In the literature, it was known that the convergence of BP over a loopy graph is not guaranteed, even though it does converge in most practical cases. Since the convergence proof for non-Gaussian input is not tractable, we will tackle this question in the next section by modifying them for Gaussian input.

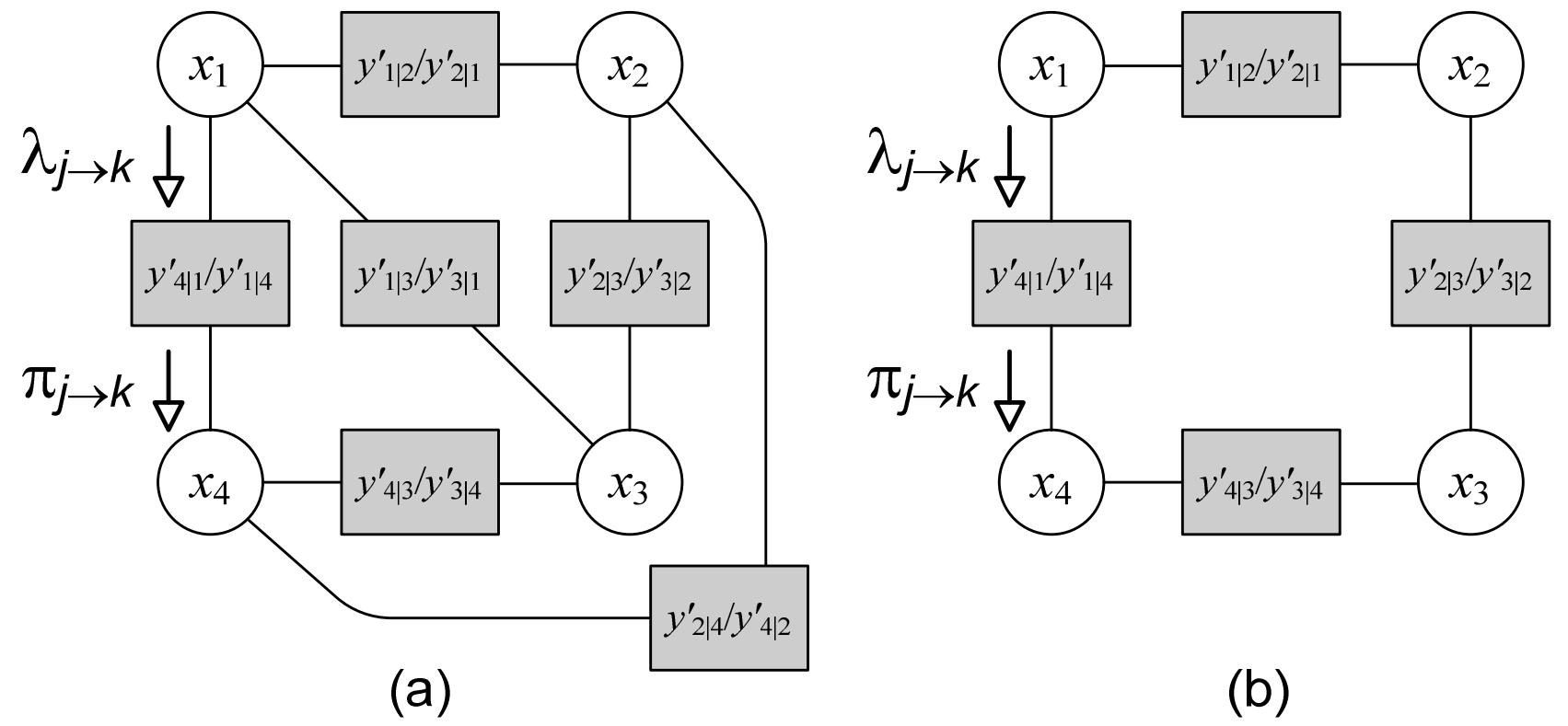

Fig.6 The Factor graph for (a) the fully-connected MRF and (b) ring-type MRF, respectively, for a 4x$N$ MIMO channel

*C. Complexity*

The computational complexity is clear from Fig.6. When compared to the MAP detector, where the marginalization in (4) is performed over all combination of ($x_1$, $x_2,\ldots,x_{j-1},x_{j+1},\ldots,x_M$) $\in\Xi^{M-1}$ for each of $2^m$ alphabet, resulting in a complexity of $O(2^{mM})$, the computational burden in the message passing rule 2 for the fully-connected MRF in Fig.6 (a) for $n$ iterations is $O(n\cdot M\cdot(M-1)\cdot 2^m)$ since the marginalization for each $M$ node is performed for $M-1$ separately of its neighbors and repeated $n$ times. Although some additional computation is required for the linear processing in (37)-(40), it is typically much smaller than $2^{m(M-1)}$, resulting in considerable computational reduction, which certainly comes from modeling through the pair-wise MRF. On the other hand, the computational complexity for the ring-type MRF in Fig.6 (b) is $O(n\cdot 2M\cdot 2^m)$, which is the same as that of message passing rule 1 and is even less than that of the fully-connected MRF.

## IV. MESSAGE PASSING WITH GAUSSIAN INPUT

In the previous section, we developed belief propagation algorithms using the pair-wise MRFs for non-Gaussian messages. The Gaussian assumption on $x_j$'s was employed to obtain the two-dimensional joint PDFs, corresponding to each edge potential, and the translation function. While, we used the exact marginalization in the message translation step. In this section, we further simplify the message passing rule by extending the Gaussian assumption to the message translation step, as was done in [16], [17], and [19], to obtain the Gaussian BP over the two MRF under consideration.

*ML detection with Gaussian Input*: With independent and identically distributed Gaussian input, $p(\mathbf{x}) = \prod_{j=1}^{M} \mathcal{CN}(x_j;0,1)$, the MAP detector in (4) becomes

$$p(x_j|\,\mathbf{y}) = A \cdot \iint \cdots \int \mathcal{CN}\left(\mathbf{y};\mathbf{Hx},\sigma^2\right)\prod_{k\neq j} \mathcal{CN}(x_k;0,1)\cdot dx_1 \cdots dx_{j-1}dx_{j+1}\cdots dx_M$$

$$= \mathcal{CN}(x_j;\mathbf{h}_j^H\mathbf{K}^{-1}\mathbf{y}, 1-\mathbf{h}_j^H\mathbf{K}^{-1}\mathbf{h}_j) \qquad (52)$$

where we appropriately chose the normalization constant, $A$, while the covariance matrix, $\mathbf{K}$, is given by

$$\mathbf{K} = \left(\mathbf{HH}^H + \sigma^2\mathbf{I}\right)$$

Noting that, in (52),

$$\mathbf{c}_j = \mathbf{K}^{-1}\mathbf{h}_j \qquad (53)$$

is the linear MMSE filter, the mean is the linear MMSE estimates of $x_j$ and the variance is the corresponding minimum MSE, i.e.,

$$\hat{x}_j = \mathbf{h}_j^H\mathbf{K}^{-1}\mathbf{y} \qquad (54)$$

$$\text{MMSE}_j = 1-\mathbf{h}_j^H\mathbf{K}^{-1}\mathbf{h}_j \qquad (55)$$

This means that linear MMSE estimation is optimum for the Gaussian input, while it generally does not hold for the non-Gaussian input.

### *A. Gaussian BP over the proposed MRF*

Assuming that $x_j$'s are Gaussian and the distributions $\lambda_{i\to j}(x_i)$, $\pi_{i\to j}(x_j)$, and $b(x_i)$ are all Gaussian PDFs, they can be characterized by their mean and variance only. This means the messages, $\lambda_{i\to j}(x_i)$ and $\pi_{i\to j}(x_j)$, and the belief, $b(x_i)$, in the message passing rules 1, 2 and 3 can be replaced with the update rule for the mean and variance pair. Since the Gaussian BP corresponding to the BP 1 over the fully connected factor graph in (9)-(11) has already been discussed in [16], here we consider only the BP 2 and 3 over the two MRF.

Let us denote the mean and variance pair of the complex Gaussian PDFs, $\pi_{i\to j}(x_j)$, $\lambda_{i\to j}(x_i)$, and $b(x_i)$ as ($\mu_{\pi,i\to j}$, $\sigma^2_{\pi,i\to j}$), ($\mu_{\lambda,i\to j}$, $\sigma^2_{\lambda,i\to j}$) and ($\mu_{b,i}$, $\sigma^2_{b,i}$). Then, the BP 2 and 3 under the Gaussian assumption can be rewritten as follows (The detailed derivations are shown in the appendix)

Gaussian BP 2G over the fully-connected MRF

```
Given the messages in the previous iteration (or the initial
messages), πk→i(xi)= (μπ,k→i, σ²π,k→i)
Compute the extrinsic information from node i to node j
```

$$\sigma^{-2}_{\lambda,i\to j} = \sum_{k\in N(i)\backslash j} \sigma^{-2}_{\pi,k\to i} \quad (56)$$

$$\mu_{\lambda,i\to j} = \frac{\sum_{k\in N(i)\backslash j} \sigma^{-2}_{\pi,k\to i}\mu_{\pi,k\to i}}{\sum_{k\in N(i)\backslash j} \sigma^{-2}_{\pi,k\to i}} \quad (57)$$

```
Translate the message λi→j(xi) to πi→j(xj)
```

$$\sigma^2_{\pi,i\to j} = \frac{1}{1+\sigma^2_{j|i}} + \frac{|a_{j|i,i}|^2}{(1+\sigma^2_{j|i})^2}\sigma^2_{\lambda,i\to j} \quad (58)$$

$$\mu_{\pi,i\to j} = \frac{y'_{j|i}}{1+\sigma^2_{j|i}} - \frac{a_{j|i,i}}{1+\sigma^2_{j|i}}\mu_{\lambda,i\to j} \quad (59)$$

```
After the iteration of (56)-(59), the belief on xi is obtained by
```

$$\sigma^{-2}_i = \sum_{k\in N(i)} \sigma^{-2}_{\pi,k\to i} \quad (60)$$

$$\mu_i = \frac{\sum_{k\in N(i)} \sigma^{-2}_{\pi,k\to i}\mu_{\pi,k\to i}}{\sum_{k\in N(i)} \sigma^{-2}_{\pi,k\to i}} \quad (61)$$

Gaussian BP 3G over the ring-type MRF (Gaussian forward-backward recursion)

```
Given the messages in the previous iteration (or the initial
messages), πk→i(xi)= (μπ,k→i, σ²π,k→i)
Extrinsic information:
```

$$\sigma^{-2}_{\lambda,i\to i\pm 1} = \sigma^{-2}_{\pi,i\mp 1\to i} \quad (62)$$

$$\mu_{\lambda,i\to i\pm 1} = \mu_{\pi,i\mp 1\to i} \quad (63)$$

```
Message translation:
```

$$\sigma^2_{\pi,i\to i\pm1} = \frac{1}{1+\sigma^2_{i\pm1|i}} + \frac{|a_{i\pm1|i,i}|^2}{(1+\sigma^2_{i\pm1|i})^2}\sigma^2_{\lambda,i\to i\pm1} \quad (64)$$

$$\mu_{\pi,i\to i\pm1} = \frac{1}{1+\sigma^2_{i\pm1|i}} y'_{i\pm1|i} - \frac{a_{i\pm1|i,i}}{1+\sigma^2_{i\pm1|i}} \mu_{\lambda,i\to i\pm1} \quad (65)$$

Final belief:

$$\sigma_i^{-2} = \sigma^{-2}_{\pi,i+1\to i} + \sigma^{-2}_{\pi,i-1\to i} \quad (66)$$

$$\mu_i = \frac{\sigma^{-2}_{\pi,i+1\to i}\mu_{\pi,i+1\to i} + \sigma^{-2}_{\pi,i-1\to i}\mu_{\pi,i+1\to i}}{\sigma^{-2}_{\pi,i+1\to i} + \sigma^{-2}_{\pi,i-1\to i}} \quad (67)$$

Especially, in the Gaussian BP 3G, we observe the following:

1) The variance and mean are updated separately (except in the final belief)
2) In (62)-(65), there are two separate message flows; one is the forward from $i$ to $i$+1 and the other is the backward from $i$ to $i$−1.
3) (63) can be combined with (65) into a single step, i.e., by plugging (63) into (65), we have

   Forward recursion: $\mu_{\pi,i\to i+1} = F_i \circ \mu_{\pi,i-1\to i}$ (68)

   Backward recursion: $\mu_{\pi,i\to i-1} = B_i \circ \mu_{\pi,i+1\to i}$ (69)

   where the operations, $F_i$ and $B_i$, are first order elementary function

   $$F_i \circ \mu \equiv u_{i+1,i} + v_{i+1,i}\cdot\mu \quad (70)$$

   $$B_i \circ \mu \equiv u_{i-1,i} + v_{i-1,i}\cdot\mu \quad (71)$$

   with

   $$u_{j,i} = \frac{y'_{j|i}}{1+\sigma^2_{j|i}} = \frac{\mathbf{h}_j^H\mathbf{K}^{-1}_{\{j,i\}}\mathbf{y}}{1+\mathbf{h}_j^H\mathbf{K}^{-1}_{\{j,i\}}\mathbf{h}_j} = \mathbf{h}_j^H\mathbf{K}^{-1}_{\{i\}}\mathbf{y} \quad (72)$$

   $$v_{j,i} = -\frac{a_{j|i,i}}{1+\sigma^2_{j|i}} = -\frac{\mathbf{h}_j^H\mathbf{K}^{-1}_{\{j,i\}}\mathbf{h}_i}{1+\mathbf{h}_j^H\mathbf{K}^{-1}_{\{j,i\}}\mathbf{h}_j} = -\mathbf{h}_j^H\mathbf{K}^{-1}_{\{i\}}\mathbf{h}_i \quad (73)$$

Here, we used (37)-(40) and, in the last, the matrix inversion lemma

$$(\mathbf{A}+\mathbf{B}\mathbf{B}^H)^{-1} = \mathbf{A}^{-1} - \mathbf{A}^{-1}\mathbf{B}(\mathbf{I}+\mathbf{B}^H\mathbf{A}^{-1}\mathbf{B})^{-1}\mathbf{B}^H\mathbf{A}^{-1}$$

4) Similar to the means, (62) can be combined with (64) into a single step, i.e.,

   Forward recursion: $\sigma^2_{\pi,i\to i+1} = F'_i \circ \sigma^2_{\pi,i-1\to i}$ (74)

   Backward recursion: $\sigma^2_{\pi,i\to i-1} = B'_i \circ \sigma^2_{\pi,i+1\to i}$ (75)

   where

$$F_i' \circ \mu \equiv u'_{i+1,i} + v'_{i+1,i} \cdot \mu \tag{76}$$

$$B_i' \circ \mu \equiv u'_{i-1,i} + v'_{i-1,i} \cdot \mu \tag{77}$$

with

$$u'_{j,i} = \frac{1}{1+\sigma_{j|i}^2} = \frac{1}{1+\mathbf{h}_j^H \mathbf{K}_{\{j,i\}}^{-1} \mathbf{h}_j} \tag{78}$$

$$v'_{j,i} = \frac{|a_{j|i,i}|^2}{(1+\sigma_{j|i}^2)^2} = \left| \frac{\mathbf{h}_j^H \mathbf{K}_{\{j,i\}}^{-1} \mathbf{h}_i}{1+\mathbf{h}_j^H \mathbf{K}_{\{j,i\}}^{-1} \mathbf{h}_j} \right|^2 = \left| \mathbf{h}_j^H \mathbf{K}_{\{i\}}^{-1} \mathbf{h}_i \right|^2 \tag{79}$$

*B. Convergence of Gaussian BP*

Regarding the convergence of Gaussian BP, it was previously shown in [24] that Gaussian BP for arbitrary topology converges to the correct mean (see also [29]), and it was shown in [16] that the Gaussian BP over the factor graph in Fig.1 (a) converges to the linear MMSE solution, even though the underlying assumption is slightly different. Based on these findings, we can conjecture that, for both the Gaussian BP of rules 2G and 3G, the mean converges to the linear MMSE solution, as also verified by the simulation in the next section. One way to prove the convergence would be to use the idea of "unwrapped tree" in [24], which, however, can be a tedious derivation in our case. Therefore, we try an alternative approach that works for GBP 3G, but not for GBP 2G. Note, however, that the derivation here is different from [16] and [17] in the underlying graphical model and the edge potentials used. The objective in this sub-section is to prove the following theorem.

*Theorem* 1: *In the Gaussian BP 3G over the ring-type MRF, the mean converges to the linear MMSE estimate (54) as the number of iteration goes to infinity.*

The proof is based on the following lemmas.

*Lemma* 2: *For arbitrary initial value* $\mu(0)$*, both the forward and backward recursion for the mean in (63) and (65) converge respectively to a unique, fixed point.*

*Proof*: Define one iteration as one complete turn of a message passing along the ring and consider, without loss of generality, the message at node 1. Based on observations 1) through 3) in the previous subsection, we obtain the recursive relations for node 1, i.e., using an arbitrary initial value $\mu(0)$, we have

$$\mu_{\pi,1\to 2}(n) = (F_M \circ \cdots F_3 \circ F_2 \circ F_1 \circ)\mu_{\pi,1\to 2}(n-1)$$
$$= (F_M \circ \cdots F_3 \circ F_2 \circ F_1 \circ)^n \mu(0) \tag{80}$$

$$\mu_{\pi,1\to M}(n) = (B_2 \circ B_3 \circ \cdots B_M \circ B_1 \circ)\mu_{\pi,1\to M}(n-1)$$
$$= (B_2 \circ B_3 \circ \cdots B_M \circ B_1 \circ)^n \mu(0) \tag{81}$$

where $n$ is the iteration number and the collective operations for one iteration of forward/backward recursion are given, respectively, by

$$F_{1,T} \circ \mu \equiv F_M \circ \cdots F_3 \circ F_2 \circ F_1 \circ \mu = f_{1,U} + f_{1,V}\mu \tag{82}$$

$$B_{1,T} \circ \mu \equiv B_2 \circ B_3 \circ \cdots B_M \circ B_1 \circ \mu = b_{1,U} + b_{1,V}\mu \tag{83}$$

for some constants, $f_{1,U}$, $f_{1,V}$, $b_{1,U}$ and $b_{1,V}$, which, in turn, are monomials of $u_{j,i}$ and $v_{j,i}$ in (72) and (73). For example, we have for $M = 4$

$$F_4 \circ F_3 \circ F_2 \circ F_1 \circ \mu = \left(u_{1,4} + v_{1,4}u_{4,3} + v_{1,4}v_{4,3}u_{3,2} + v_{1,4}v_{4,3}v_{3,2}u_{2,1}\right) + \left(v_{1,4}v_{4,3}v_{3,2}v_{2,1}\right)\cdot \mu$$

$$B_2 \circ B_3 \circ B_4 \circ B_1 \circ \mu = \left(u_{1,2} + v_{1,2}u_{2,3} + v_{1,2}v_{2,3}u_{3,4} + v_{1,2}v_{2,3}v_{3,4}u_{4,1}\right) + \left(v_{1,2}v_{2,3}v_{3,4}v_{4,1}\right)\cdot \mu$$

for which

$$f_{1,U} = u_{1,4} + v_{1,4}u_{4,3} + v_{1,4}v_{4,3}u_{3,2} + v_{1,4}v_{4,3}v_{3,2}u_{2,1}$$

$$f_{1,V} = v_{1,4}v_{4,3}v_{3,2}v_{2,1}\,.$$

$$b_{1,U} = u_{1,2} + v_{1,2}u_{2,3} + v_{1,2}v_{2,3}u_{3,4} + v_{1,2}v_{2,3}v_{3,4}u_{4,1}$$

$$b_{1,V} = v_{1,2}v_{2,3}v_{3,4}v_{4,1}\,.$$

Here, we can show that $f_{1,V}$ and $b_{1,V}$ are given respectively by

$$f_{1,V} = \prod_{j=1}^{M} v_{j,j-1} \quad \text{and} \quad b_{1,V} = \prod_{j=1}^{M} v_{j,j+1} \tag{84}$$

On the other hand, using (82) and (83), (80) and (81) become

$$\mu_{\pi,1\to 2}(n) = (F_{1,T}\circ)^n \mu(0) = f_{1,U}\cdot \sum_{k=0}^{n-1} f_{1,V}^k + f_{1,V}^n \cdot \mu(0) \tag{85}$$

$$\mu_{\pi,1\to M}(n) = (B_{1,T}\circ)^n \mu(0) = b_{1,U}\cdot \sum_{k=0}^{n-1} b_{1,V}^k + b_{1,V}^n \cdot \mu(0) \tag{86}$$

where, from the fact to be proved in the next lemma that $|f_{i,V}| < 1$ and $|b_{i,V}| < 1$, we have

$$f_{1,V}^n \cdot \mu(0) \to 0,\ b_{1,V}^n \cdot \mu(0) \to 0 \ \text{ as } n\to\infty$$

Therefore, the unique fixed point of the mean in GBP 3G is given by

$$\lim_{n\to\infty} \mu_{\pi,1\to 2}(n) \to f_{1,U}\cdot \sum_{k=0}^{\infty} f_{1,V}^k = \frac{f_{1,U}}{1 - f_{1,V}} \tag{87}$$

$$\lim_{n\to\infty} \mu_{\pi,1\to M}(n) \to b_{1,U}\cdot \sum_{k=0}^{\infty} b_{1,V}^k = \frac{b_{1,U}}{1 - b_{1,V}} \tag{88}$$

■

*lemma* 3: $|f_{i,V}| = \left|\prod_{j=1}^{M} v_{j,j-1}\right| < 1$ and $|b_{i,V}| = \left|\prod_{j=1}^{M} v_{j,j+1}\right| < 1$ for all $i$.

*Proof*: By plugging (73) into (84), we have for any $i$

$$
\begin{aligned}
\left|f_{i,V}\right| &= \left|\prod_{j=1}^{M} \frac{\mathbf{h}_j^H \mathbf{K}_{\{j,j-1\}}^{-1} \mathbf{h}_{j-1}}{1+\mathbf{h}_j^H \mathbf{K}_{\{j,j-1\}}^{-1} \mathbf{h}_j}\right| \\
&= \left|\prod_{j=1}^{M} \mathbf{h}_j^H \mathbf{K}_{\{j-1\}}^{-1} \mathbf{h}_{j-1}\right| \\
&= \left|\mathbf{h}_1^H \mathbf{K}_{\{M\}}^{-1} \mathbf{h}_M \mathbf{h}_M^H \mathbf{K}_{\{M-1\}}^{-1} \cdots \mathbf{K}_{\{2\}}^{-1} \mathbf{h}_2 \mathbf{h}_2^H \mathbf{K}_{\{1\}}^{-1} \mathbf{h}_1\right| \\
&= \left|\operatorname{tr}\left(\mathbf{h}_1 \mathbf{h}_1^H \mathbf{K}_{\{M\}}^{-1} \mathbf{h}_M \mathbf{h}_M^H \mathbf{K}_{\{M-1\}}^{-1} \cdots \mathbf{K}_{\{2\}}^{-1} \mathbf{h}_2 \mathbf{h}_2^H \mathbf{K}_{\{1\}}^{-1}\right)\right| \\
&\le \left|\prod_{j=1}^{M} \operatorname{tr}\left(\mathbf{h}_j \mathbf{h}_j^H \mathbf{K}_{\{j-1\}}^{-1}\right)\right| \\
&= \left|\prod_{j=1}^{M} \mathbf{h}_j^H \mathbf{K}_{\{j-1\}}^{-1} \mathbf{h}_j\right| \\
&= \prod_{j=1}^{M} \frac{\mathbf{h}_j^H \mathbf{K}_{\{j,j-1\}}^{-1} \mathbf{h}_j}{1+\mathbf{h}_j^H \mathbf{K}_{\{j,j-1\}}^{-1} \mathbf{h}_j} \\
&= \prod_{j=1}^{M} \frac{\sigma_{j|j-1}^2}{1+\sigma_{j|j-1}^2} < 1
\end{aligned}
$$

where, in the third line, we used $\mathbf{a}^H \mathbf{b} = \operatorname{tr}(\mathbf{b}\mathbf{a}^H)$ for arbitrary vector **a** and **b**, and, in the fourth line, $\operatorname{tr}(\mathbf{AB}) \le \operatorname{tr}(\mathbf{A})\operatorname{tr}(\mathbf{B})$ for arbitrary non-negative definite matrices **A** and **B**. In the third line from the bottom, we used the matrix inversion lemma to get

$$
\begin{aligned}
\mathbf{h}_j^H \mathbf{K}_{\{j+1\}}^{-1} &= \mathbf{h}_j^H \left(\mathbf{K}_{\{j,j+1\}} + \mathbf{h}_j^H\right)^{-1} \\
&= \mathbf{h}_j^H \left(\mathbf{K}_{\{j,j+1\}}^{-1} - \mathbf{K}_{\{j,j+1\}}^{-1} \mathbf{h}_j (1+\mathbf{h}_j^H \mathbf{K}_{\{j,j+1\}}^{-1} \mathbf{h}_j)^{-1} \mathbf{h}_j^H \mathbf{K}_{\{j,j+1\}}^{-1}\right) \\
&= \left(1 - \frac{\mathbf{h}_j^H \mathbf{K}_{\{j,j+1\}}^{-1} \mathbf{h}_j}{1+\mathbf{h}_j^H \mathbf{K}_{\{j,j+1\}}^{-1} \mathbf{h}_j}\right) \mathbf{h}_j^H \mathbf{K}_{\{j,j+1\}}^{-1} \\
&= \left(\frac{1}{1+\mathbf{h}_j^H \mathbf{K}_{\{j,j+1\}}^{-1} \mathbf{h}_j}\right) \mathbf{h}_j^H \mathbf{K}_{\{j,j+1\}}^{-1}
\end{aligned}
$$

For the backward recursion, $|b_{i,V}| < 1$ can also be proved in a similar way. ■

In (85) and (86), we see that the convergence rate depends on

$$
\left|\prod_{j=1}^{M} \mathbf{h}_j^H \mathbf{K}_{\{j-1\}}^{-1} \mathbf{h}_{j-1}\right| \le \prod_{j=1}^{M} \frac{\sigma_{j|j-1}^2}{1+\sigma_{j|j-1}^2} < 1
$$

which is similar to the result in [16]. Note that $\mathbf{h}_j^H \mathbf{K}_{\{j-1\}}^{-1} \mathbf{h}_{j-1}$ reflects the channel correlation between neighbor antennas.

On the other hand, the operations, $F_i$ and $B_i$, are not permutable such that $F_i \circ F_j \circ \mu$ and $F_j \circ F_i \circ \mu$ may not be the same, and so are $F_{j,T} \circ \mu$ and $F_{i,T} \circ \mu$ for $j \neq i$. That is, the fixed point for each node may be different from one another. Note also that the fixed point of the mean, (85) and (86), can be obtained by the direct computation of $f_{j,U}$, $f_{j,V}$, $b_{j,U}$ and $b_{j,V}$ using (72) and (73). However, it is not tractable for large $M$. Instead of direct computation, we can obtain the fixed point of the mean using the following lemma.

*Lemma* 4: *In the forward recursion,* $\mu_{\pi,i\to i+1}(n)$ *is the linear MMSE estimates of* $x_{i+1}$ *provided that the previous message,* $\mu_{\pi,i-1\to i}(n)$*, is the linear MMSE estimates of* $x_i$*. Likewise, in the backward recursion,* $\mu_{\pi,i\to i-1}(n)$ *is the linear MMSE estimates of* $x_{i-1}$ *provided that* $\mu_{\pi,i+1\to i}(n)$ *is the linear MMSE estimates of* $x_i$*.*

*Proof*: From (53), the linear MMSE estimate of $x_i$ is given by $\mathbf{h}_i^H \mathbf{K}^{-1}\mathbf{y}$. And, hence, the proof is to show from (68) and (69) that

$$\mathbf{h}_{i+1}^H \mathbf{K}^{-1}\mathbf{y} = F_i \circ (\mathbf{h}_i^H \mathbf{K}^{-1}\mathbf{y}) = u_{i+1,i} + v_{i+1,i} \cdot (\mathbf{h}_i^H \mathbf{K}^{-1}\mathbf{y}) \tag{89}$$

$$\mathbf{h}_{i-1}^H \mathbf{K}^{-1}\mathbf{y} = B_i \circ (\mathbf{h}_i^H \mathbf{K}^{-1}\mathbf{y}) = u_{i-1,i} + v_{i-1,i} \cdot (\mathbf{h}_i^H \mathbf{K}^{-1}\mathbf{y}) \tag{90}$$

where $u_{j,i}$ and $v_{j,i}$ are given by (72) and (73). Plugging these into the r.h.s. of (89) for the forward recursion, we finally have

$$\begin{aligned} u_{i+1,i} + v_{i+1,i} \cdot (\mathbf{h}_i^H \mathbf{K}^{-1}\mathbf{y}) &= \mathbf{h}_{i+1}^H \mathbf{K}_{\{i\}}^{-1}\mathbf{y} - \mathbf{h}_{i+1}^H \mathbf{K}_{\{i\}}^{-1}\mathbf{h}_i \mathbf{h}_i^H \mathbf{K}^{-1}\mathbf{y} \\ &= \mathbf{h}_{i+1}^H \left(\mathbf{K}_{\{i\}}^{-1} - \mathbf{K}_{\{i\}}^{-1}\mathbf{h}_i \mathbf{h}_i^H \mathbf{K}^{-1}\right)\mathbf{y} \\ &= \mathbf{h}_{i+1}^H \left(\mathbf{K}_{\{i\}}^{-1} - \mathbf{K}_{\{i\}}^{-1}\mathbf{h}_i \frac{\mathbf{h}_i^H \mathbf{K}_{\{i\}}^{-1}}{1 + \mathbf{h}_i^H \mathbf{K}_{\{i\}}^{-1}\mathbf{h}_i}\right)\mathbf{y} \\ &= \mathbf{h}_{i+1}^H \mathbf{K}^{-1}\mathbf{y} \end{aligned}$$

Similarly, for the backward recursion, we obtain

$$u_{i-1,i} + v_{i-1,i} \cdot (\mathbf{h}_i^H \mathbf{K}^{-1}\mathbf{y}) = \mathbf{h}_{i-1}^H \mathbf{K}^{-1}\mathbf{y}$$ ■

*Proof of theorem* 1: Theorem 1 is now obvious from the above lemmas, i.e., from lemma 2 the mean in the Gaussian BP over the ring-type MRF converges to a fixed point and lemma 4 shows that, once converged, the fixed point (mean) is given by the linear MMSE estimates. ■

Since the message update rule for the variance in (74) and (75) have the same form as in (68) and (69), we also can prove the convergence of the variance in GBP 3G, which can be summarized by the following lemma.

*Lemma* 5: *For arbitrary initial value* $\sigma^2(0)$, *both the forward and backward recursion for the variance in (62) and (64) converges respectively to a unique, fixed point.*

The proof is similar to that of lemma 3. Unfortunately, however, the fixed point is not necessarily correct, i.e., may not equal to the MMSE in (55), as also confirmed in [24]. In [28], the convergence property of the BP over such ring-type graph was shown to be optimal for binary input. For Gaussian input, however, it is optimal only in the mean, and we cannot say so in a strict sense.

## V. SIMULATION RESULTS

In this section, we present the simulation results for the iterative algorithms with DVB-S2 LDPC code of rates 3/4 and length 64800 [30]. The performances of ML, MMSE and the bi-diagonalization approach in [9] are also evaluated as references. We generated a block of independent and identically distributed MIMO channel matrix, of which each element is also i.i.d. complex Gaussian random variable with mean 0 and variance 1. The resulting channel can be regarded as a fully interleaved frequency selective MIMO channel that can be seen on top of the orthogonal frequency division multiplexing (OFDM), especially for those channels where the transmission bandwidth is much larger than the channel coherence bandwidth.

Fig.7 shows a comparison of bit error rate performance as a function of SNR ($1/\sigma^2$) for ML, MMSE, the bi-diagonalization approach in [9] and the proposed MRF-based detector of fully-connected and ring type. We used a 4×4 antenna configuration and QPSK modulation. Fig.7 shows that the pair-wise MRF based detector performs as well as ML. The SNR gap between the proposed scheme and the ML is shown to be around 0.1 and 0.3, respectively. In Fig.7, the number of iterations was set to 4 for BP 3 over ring-type MRF and 3 for BP 2 over fully connected MRF. These numbers of iterations are based on the simulation in Fig.8 and 9 for the ring-type MRF and fully-connected MRF, respectively.

From Fig.8 and 9, the BP 3 over the ring-type MRF is shown to converge in 4 iterations for the current antenna configurations while the BP 2 over the fully connected MRF is shown to converge in 3 iterations and, with more number of iteration, the performance is slightly degraded. The difference in the convergence rate seems to stem from the short cycles in the graphical models. The factor graphs in Fig.6 (a) for the fully connected MRF is more densely connected than the one in Fig.6 (b) for the ring-type MRF. In densely connected graph, the message will propagate faster than in sparsely connected graph. Such situation can also be observed in the Gaussian BP. On the other hand, in densely connected graph, the messages may circulate along the short loops preventing steady convergence. The slight degradation with more iteration than 3 in Fig.8 seem to be due to this reason.

Fig.10 shows the convergence behavior of the Gaussian BP discussed in the previous section. We generated 20 independent MIMO channels of 4×4 and performed simulations for each channel to obtain the residual MSE normalized by the minimum MSE in (55) as a function of the iteration number, i.e.,

$$e(n) \equiv \frac{|\boldsymbol{\mu}(n) - \mathbf{x}|^2}{\sum_j \mathrm{MMSE}_j}$$

The Fig.10 shows that both Gaussian BP 2G over fully-connected MRF and 3G over ring-type MRF finally converge to the MMSE solution. In the convergence rate, however, Gaussian BP 2G is shown to be faster than that of 3G on the average. This is one of the expected results. As can be inferred from the proof of lemma 3, the convergence of the GBP depends on the correlation between neighbor antennas. In ring-type MRF, there is only one message pathway and the convergence will get slower if antennas are badly ordered. In fully-connected MRF, however, there exist other pathways (loops) which seem to facilitate the message passing even if some neighbor pairs have high antenna correlation.

## VI. CONCLUDING REMARKS

In this paper, low complexity, iterative MIMO detection algorithms were derived based on the pair-wise Markov random fields (MRF) with the potential functions that are obtained by marginalizing the posterior joint probability density under the Gaussian assumption. We examined two MRFs, namely, the fully-connected and ring-type pair-wise MRF, where the latter is shown to be an extension of the previous work in [9] and [10].

The factor graphs corresponding to the two MRF are rather sparse in the sense that the number of edges connected to a factor node, i.e., edge degree, is only 2 and, thus, the message passing becomes much easier than that over the fully connected factor graph. The simulation results show that the proposed schemes perform very close to the ML detection.

We also investigated the proposed algorithm for Gaussian input, where it was shown that,

for the Gaussian BP over the fully-connected and ring-type MRF, the mean converges to the linear MMSE estimates, even though the variance does not converge to the correct value. These results lie on the same line of those in [16], [17], [24] and [29]. Compared to Gaussian BP over the ring-type MRF, that over the fully-connected MRF shows faster convergence rate.

Although the convergence of the Gaussian BPs over the two MRFs are shown to be guaranteed, it seems not for non-Gaussian message as the performance of BP 2 and 3 for non-Gaussian case are slightly degraded with more than 3 and 4 iterations, respectively. This phenomenon might stem from the short cycles in their graphical model and may be avoided by utilizing "global iteration" between MIMO detection and channel decoding. That is, by employing an appropriate channel code and interleaver, message circulation along local short cycles can be broken up not only for steady convergence but also for better performance. We leave this for our future work.

## APPENDIX A: DETAILED DERIVATIONS OF (44) AND THE GAUSSIAN BP

To derive (44) and the Gaussian BP rule, (56)-(61), we use the properties of the Gaussian PDF in [21], some of which are as follows

1) $\mathcal{CN}(x;\mu,\sigma^2)=\mathcal{CN}(\mu;x,\sigma^2)=\mathcal{CN}(x-\mu;0,\sigma^2)=\mathcal{CN}(\mu-x;0,\sigma^2)$ (A.1)

2) $\mathcal{CN}(ax+b;\mu,\sigma^2)=\mathcal{CN}\left(x;\frac{\mu-b}{a},\frac{\sigma^2}{|a|^2}\right)$ (A.2)

3) $\mathcal{CN}(x;\mu_1,\sigma_1^2)\cdot\mathcal{CN}(x;\mu_2,\sigma_2^2)$

$$=\mathcal{CN}\left(x;\frac{\sigma_1^{-2}\mu_1+\sigma_2^{-2}\mu_2}{\sigma_1^{-2}+\sigma_2^{-2}},\frac{1}{\sigma_1^{-2}+\sigma_2^{-2}}\right)\cdot\mathcal{CN}(\mu_1;\mu_2,\sigma_1^2+\sigma_1^2) \quad \text{(A.3)}$$

4) $\int\mathcal{CN}(x;\mu_1,\sigma_1^2)\cdot\mathcal{CN}(x;\mu_2,\sigma_2^2)\cdot dx=\mathcal{CN}(\mu_1;\mu_2,\sigma_1^2+\sigma_1^2)$ (A.4)

Using these, (44) is obtained by direct computation as follows.

$$\begin{aligned}
p(x_j|x_i,y'_{j|i}) &= \frac{\mathcal{CN}(y'_{j|i};a_{j|i,j}x_j+a_{j|i,i}x_i,\sigma_{j|i}^2)\cdot\mathcal{CN}(x_j;0,1)}{\mathcal{CN}(y'_{j|i};a_{j|i,i}x_i,\sigma_{j|i}^2+|a_{j|i,j}|^2)} \\
&= \frac{\mathcal{CN}(a_{j|i,j}x_j;y'_{j|i}-a_{j|i,i}x_i,\sigma_{j|i}^2)\cdot\mathcal{CN}(x_j;0,1)}{\mathcal{CN}(y'_{j|i};a_{j|i,i}x_i,\sigma_{j|i}^2+|a_{j|i,j}|^2)} \\
&= \frac{\mathcal{CN}\left(x_j;\frac{1}{a_{j|i,j}}(y'_{j|i}-a_{j|i,i}x_i),\frac{\sigma_{j|i}^2}{|a_{j|i,j}|^2}\right)\cdot\mathcal{CN}(x_j;0,1)}{\mathcal{CN}(y'_{j|i};a_{j|i,i}x_i,\sigma_{j|i}^2+|a_{j|i,j}|^2)}
\end{aligned}$$

$$= \mathcal{CN}\left( x_j; \frac{\frac{|a_{j|i,j}|^2}{a_{j|i,j}\sigma_{j|i}^2}\left(y'_{j|i} - a_{j|i,i}x_i\right)}{1+\left(\sigma_{j|i}^2 / |a_{j|i,j}|^2\right)^{-1}}, \left(\left(\frac{\sigma_{j|i}^2}{|a_{j|i,j}|^2}\right)^{-1} + 1\right)^{-1} \right)$$

$$\cdot \frac{\mathcal{CN}\left( \frac{1}{a_{j|i,j}}\left(y'_{j|i} - a_{j|i,i}x_i\right); 0, 1+\frac{\sigma_{j|i}^2}{|a_{j|i,j}|^2}\right)}{\mathcal{CN}(y'_{j|i}; a_{j|i,i}x_i, \sigma_{j|i}^2 + |a_{j|i,j}|^2)}$$

$$= \mathcal{CN}\left( x_j; \frac{\frac{1}{a_{j|i,j}}\left(y'_{j|i} - a_{j|i,i}x_i\right)}{\frac{\sigma_{j|i}^2}{|a_{j|i,j}|^2}+1}, \frac{\sigma_{j|i}^2}{|a_{j|i,j}|^2}\cdot\left(1+\frac{\sigma_{j|i}^2}{|a_{j|i,j}|^2}\right)^{-1} \right)$$

$$\cdot \frac{\mathcal{CN}\left(y'_{j|i} - a_{j|i,i}x_i; 0, |a_{j|i,j}|^2 + \sigma_{j|i}^2\right)}{\mathcal{CN}(y'_{j|i}; a_{j|i,i}x_i, \sigma_{j|i}^2 + |a_{j|i,j}|^2)}$$

$$= \mathcal{CN}\left( x_j; \frac{a_{j|i,j}^*}{\sigma_{j|i}^2 + |a_{j|i,j}|^2}\left(y'_{j|i} - a_{j|i,i}x_i\right), \frac{\sigma_{j|i}^2}{\sigma_{j|i}^2 + |a_{j|i,j}|^2} \right) \quad \text{(A.5)}$$

On the other hand, assuming the Gaussian messages,

$$\pi_{k\to i}(x_i) = \mathcal{CN}\left(x_i; \mu_{\pi,k\to i}, \sigma_{\pi,k\to i}^2\right) \quad \text{(A.6)}$$

$$\lambda_{i\to j}(x_i) = \mathcal{CN}\left(x_i; \mu_{\lambda,i\to j}, \sigma_{\lambda,i\to j}^2\right) \quad \text{(A.7)}$$

we have for the extrinsic information computation in (45)

$$\begin{aligned}
\lambda_{i\to j}(x_i) &= \prod_{k\in N(i)\setminus j} \pi_{k\to i}(x_i) \\
&= \prod_{k\in N(i)\setminus j} \mathcal{CN}\left(x_i; \mu_{\pi,k\to i}, \sigma_{\pi,k\to i}^2\right) \\
&\propto \mathcal{CN}\left( x_i; \frac{\sum_{k\in N(i)\setminus j} \sigma_{\pi,k\to i}^{-2}\mu_{\pi,k\to i}}{\sum_{k\in N(i)\setminus j} \sigma_{\pi,k\to i}^{-2}}, \left(\sum_{k\in N(i)\setminus j} \sigma_{\pi,k\to i}^{-2}\right)^{-1} \right) \\
&= \mathcal{CN}\left(x_i; \mu_{\lambda,i\to j}, \sigma_{\lambda,i\to j}^2\right)
\end{aligned} \quad \text{(A.8)}$$

and, similarly, for the belief update

$$\begin{aligned}
b(x_i) &= \prod_{k\in N(i)} \pi_{k\to i}(x_i) \\
&= \prod_{k\in N(i)} \mathcal{CN}\left(x_i; \mu_{\pi,k\to i}, \sigma_{\pi,k\to i}^2\right) \\
&\propto \mathcal{CN}\left( x_i; \frac{\sum_{k\in N(i)} \sigma_{\pi,k\to i}^{-2}\mu_{\pi,k\to i}}{\sum_{k\in N(i)} \sigma_{\pi,k\to i}^{-2}}, \left(\sum_{k\in N(i)} \sigma_{\pi,k\to i}^{-2}\right)^{-1} \right) \\
&= \mathcal{CN}\left(x_i; \mu_{\lambda,i\to j}(n+1), \sigma_{\lambda,i\to j}^2(n+1)\right)
\end{aligned} \quad \text{(A.9)}$$

In (A.8) and (A.9), we can identify the message update rules, (56), (57), (60) and (61).

For the message translation in (46), we first rewrite the last line of (A.5) as

$$p(x_j | x_i, y'_{j|i}) = \mathcal{CN}\left( x_j; \frac{a^*_{j|i,j}}{\sigma^2_{j|i} + |a_{j|i,j}|^2} \left( y'_{j|i} - a_{j|i,i} x_i \right), \frac{\sigma^2_{j|i}}{\sigma^2_{j|i} + |a_{j|i,j}|^2} \right)$$
$$= \mathcal{CN}\left( \frac{\sigma^2_{j|i} + |a_{j|i,j}|^2}{a^*_{j|i,j}} x_j; \left( y'_{j|i} - a_{j|i,i} x_i \right), \frac{\sigma^2_{j|i} (\sigma^2_{j|i} + |a_{j|i,j}|^2)}{|a_{j|i,j}|^2} \right)$$
$$= \mathcal{CN}\left( x_i; \frac{1}{a_{j|i,i}} \left( y'_{j|i} - \frac{\sigma^2_{j|i} + |a_{j|i,j}|^2}{a^*_{j|i,j}} x_j \right), \frac{\sigma^2_{j|i} (\sigma^2_{j|i} + |a_{j|i,j}|^2)}{|a_{j|i,j}|^2 |a_{j|i,i}|^2} \right)$$

By plugging the last equation into (46) and assuming the Gaussian messages, (A.6) and (A.7), we have

$$\pi_{i\to j}(x_j) = \int_{x_i} p(x_j \mid x_i, y'_{ji}) \cdot \lambda_{i\to j}(x_i) \cdot dx_i$$
$$= \int_{x_i} \mathcal{CN}\left( x_i; \frac{1}{a_{j|i,i}} \left( y'_{j|i} - \frac{\sigma^2_{j|i} + |a_{j|i,j}|^2}{a^*_{j|i,j}} x_j \right), \frac{\sigma^2_{j|i} (\sigma^2_{j|i} + |a_{j|i,j}|^2)}{|a_{j|i,j}|^2 |a_{j|i,i}|^2} \right) \cdot \mathcal{CN}\left( x_i; \mu_{\lambda,i\to j}, \sigma^2_{\lambda,i\to j} \right) \cdot dx_i$$
$$= \mathcal{CN}\left( \frac{1}{a_{j|i,i}} \left( y'_{j|i} - \frac{\sigma^2_{j|i} + |a_{j|i,j}|^2}{a^*_{j|i,j}} x_j \right); \mu_{\lambda,i\to j}, \frac{\sigma^2_{j|i} (\sigma^2_{j|i} + |a_{j|i,j}|^2)}{|a_{j|i,j}|^2 |a_{j|i,i}|^2} + \sigma^2_{\lambda,i\to j} \right)$$
$$= \mathcal{CN}\left( y'_{j|i} - \frac{\sigma^2_{j|i} + |a_{j|i,j}|^2}{a^*_{j|i,j}} x_j - a_{j|i,i} \mu_{\lambda,i\to j}; 0, \frac{\sigma^2_{j|i} (\sigma^2_{j|i} + |a_{j|i,j}|^2)}{|a_{j|i,j}|^2} + |a_{j|i,i}|^2 \sigma^2_{\lambda,i\to j} \right)$$
$$= \mathcal{CN}\left( x_j; \frac{a^*_{j|i,j}}{\sigma^2_{j|i} + |a_{j|i,j}|^2} \left( y'_{j|i} - a_{j|i,i} \mu_{\lambda,i\to j} \right), \frac{\sigma^2_{j|i}}{\sigma^2_{j|i} + |a_{j|i,j}|^2} + \frac{|a_{j|i,j}|^2}{(\sigma^2_{j|i} + |a_{j|i,j}|^2)^2} |a_{j|i,i}|^2 \sigma^2_{\lambda,i\to j} \right)$$
$$= \mathcal{CN}\left( x_j; \mu_{\pi,i\to j}, \sigma^2_{\pi,i\to j} \right) \tag{A.10}$$

By comparing the mean and variance in the last two lines, we obtain the message passing rules of (58) and (59), respectively.

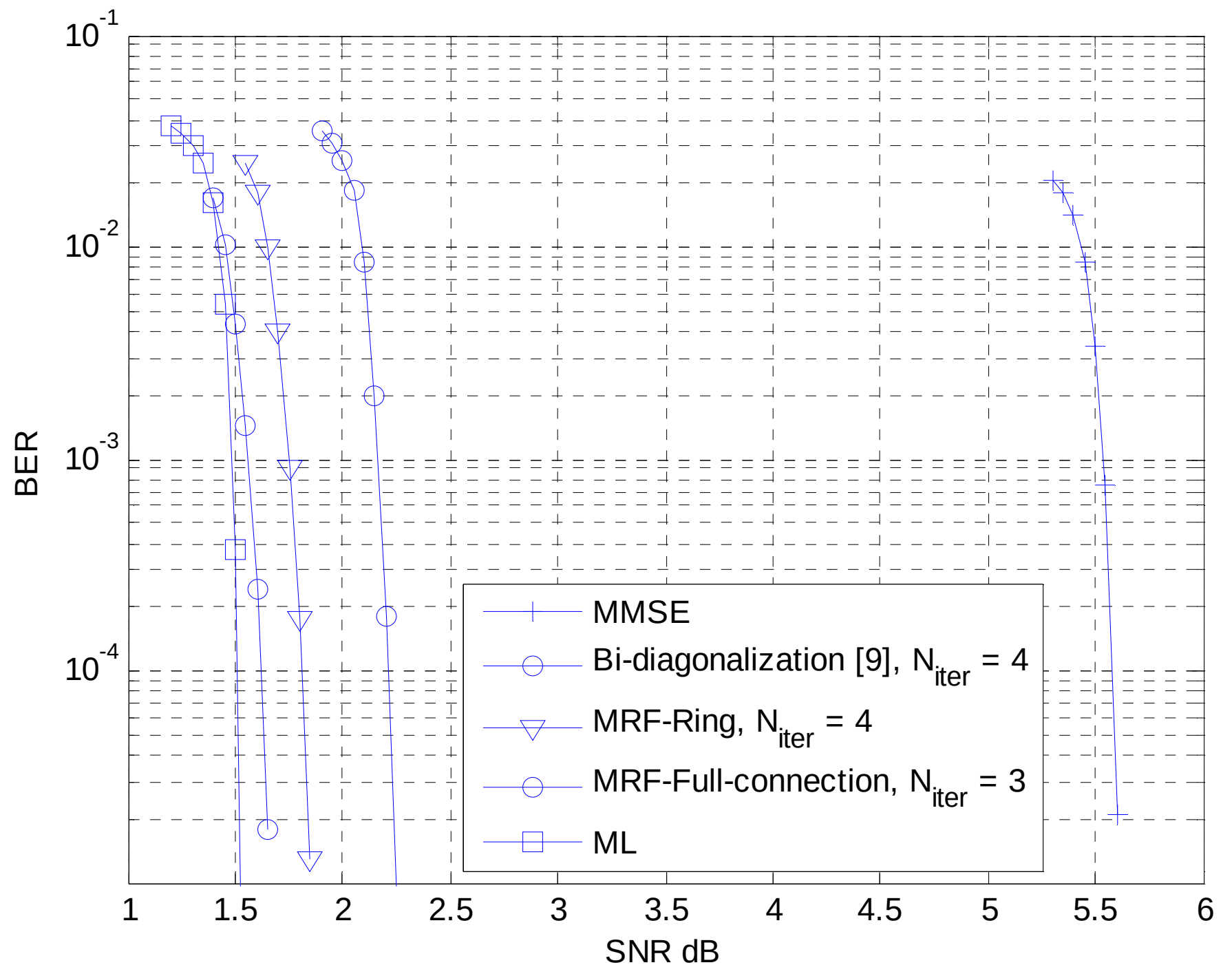

Fig.7 A comparison of Bit error rate performance of MMSE, MAP and the proposed detectors as a function of SNR ($1/\sigma^2$); 4×4 antenna configuration, QPSK modulation with DVB-S2 LDPC code of rate 3/4 (length 64800)

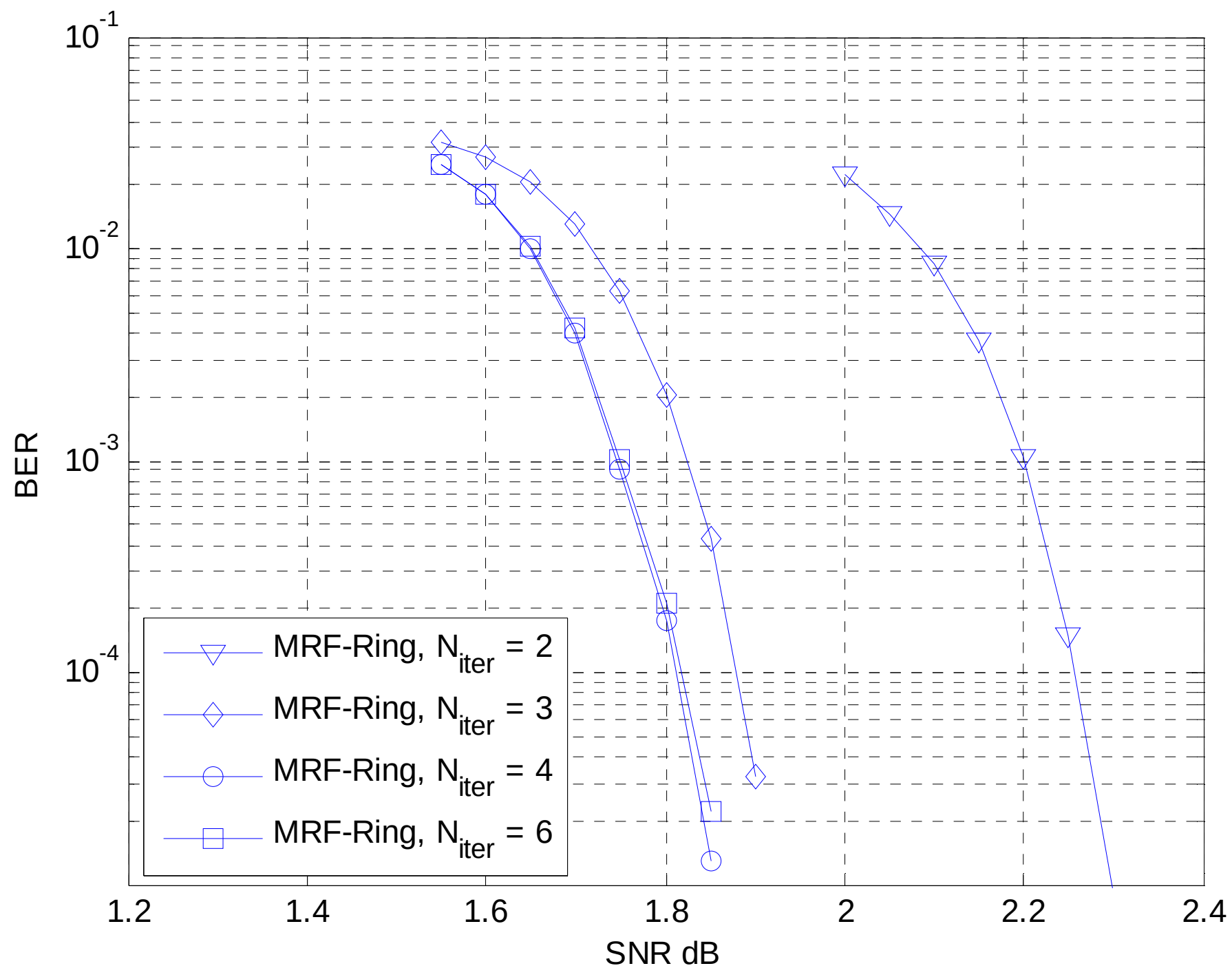

Fig.8 Bit error rate performance of BP 2 over ring-type MRF as a function of SNR ($1/\sigma^2$); 4×4 antenna configuration, QPSK modulation with DVB-S2 LDPC code of rate 3/4 (length 64800), the number of iterations = 2, 3, 4 and 6, respectively

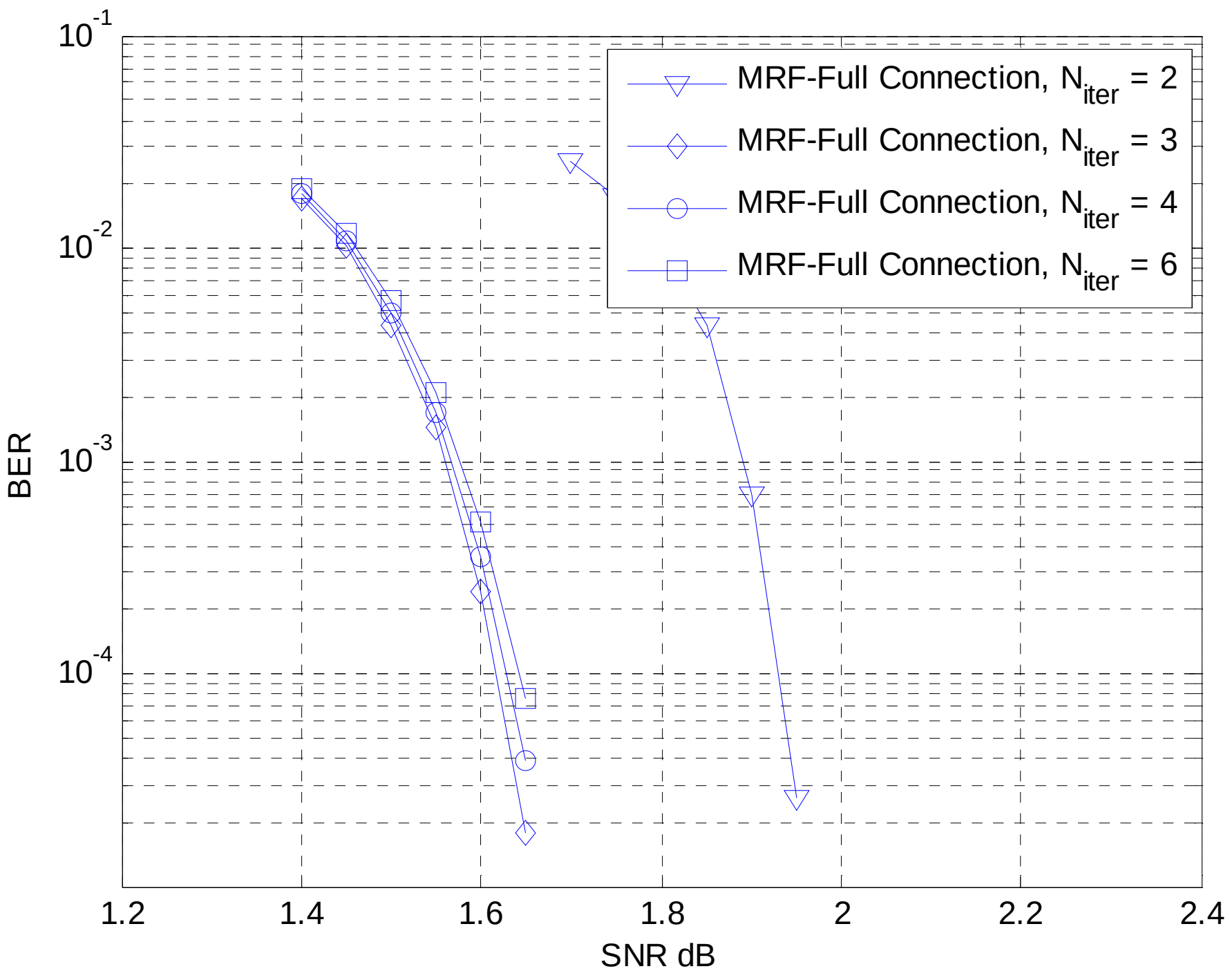

Fig.9 Bit error rate performance of BP 1 over fully connected MRF as a function of SNR ($1/\sigma^2$); 4×4 antenna configuration, QPSK modulation with DVB-S2 LDPC code of rate 3/4 (length 64800), the number of iterations = 2, 3, 4 and 6, respectively

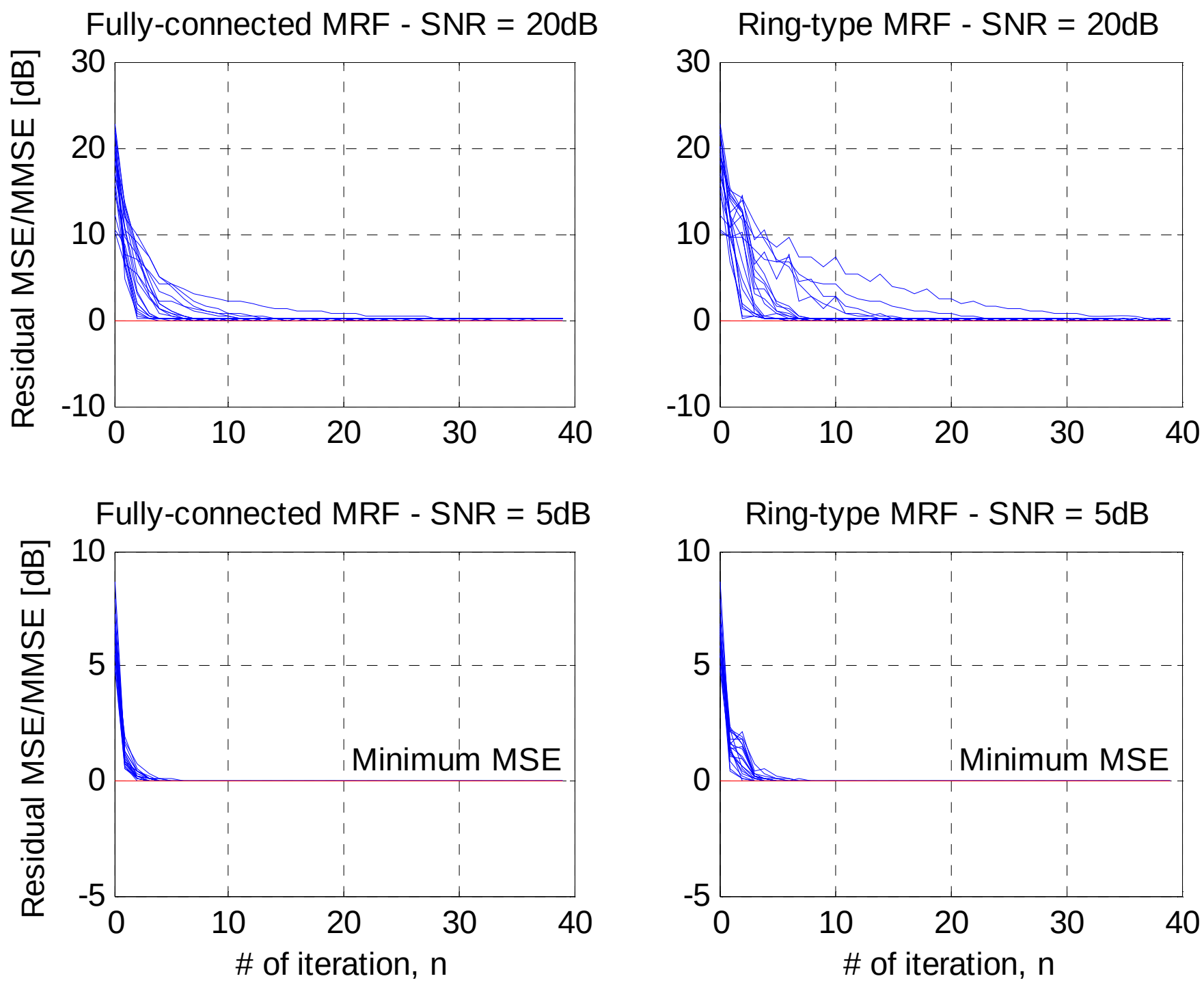

Fig.10 Convergence characteristics of the Gaussian BP over the fully-connected and ring-type MRF, respectively; 4×4 antenna configuration, SNR = 5, 20 dB